\documentclass[referee]{sn-jnl}
\usepackage[super,sort&compress,comma]{natbib}
\usepackage{multibib}
\newcites{methods}{Methods References}

\usepackage{graphicx}%
\usepackage{multirow}%
\usepackage{amsmath,amssymb,amsfonts}%
\usepackage{amsthm}%
\usepackage{mathrsfs}%
\usepackage[title]{appendix}%
\usepackage{xcolor}%
\usepackage{textcomp}%
\usepackage{manyfoot}%
\usepackage{booktabs}%
\usepackage{algorithm}%
\usepackage{algorithmicx}%
\usepackage{algpseudocode}%
\usepackage{listings}%
\usepackage{setspace}
\usepackage{amsmath}
\usepackage{times}
\usepackage{float}
\usepackage{bm}

\usepackage{xspace}

\usepackage{caption}
\DeclareCaptionLabelSeparator{bar}{\xspace\rule{0.4pt}{1em}\xspace}
\DeclareCaptionType{extfigure}[Extended Data Fig.][List of Extended Data Figures]
\makeatletter
\renewcommand\@biblabel[1]{#1.}
\makeatother

\newcommand{\etah}{\ensuremath{\eta_{xzyz}}\xspace}
\newcommand{\rcl}{\ensuremath{\alpha}-RuCl\ensuremath{_3}\xspace}

\newcommand{\asym}[2]{\varepsilon_{#1}\dot{\varepsilon}_{#2} - \dot{\varepsilon}_{#1}\varepsilon_{#2}}

\renewcommand{\figurename}{Fig.}
\renewcommand{\refname}{References and Notes}

\newenvironment{natureabstract}{%
  \par\vspace{0.25\baselineskip}%
  \noindent\begingroup
  \doublespacing            % double line spacing
  \bfseries                 % boldface text
  \small                    % (optional: adjust size if needed)
  {\large\bfseries Abstract}\quad    % bold heading
}{%
  \par\endgroup\vspace{0.75\baselineskip}%
}

\unnumbered

\begin{document}

%----------------------------------------------------
% TITLE AUTHORS AND ABSTRACT
%----------------------------------------------------

\title{Phonon Hall Viscosity and the Intrinsic Thermal Hall Effect of \rcl}

\author[1]{\fnm{Avi} \sur{Shragai}}\nomail
\author[2]{\fnm{Ezekiel} \sur{Horsley}}\nomail
\author[2]{\fnm{Subin} \sur{Kim}}\nomail
\author[2]{\fnm{Young-June} \sur{Kim}}\nomail
\author*[1,3]{\fnm{B.J.} \sur{Ramshaw}}\email{bradramshaw@cornell.edu}

\affil*[1]{\orgdiv{Laboratory of Atomic and Solid State Physics}, \orgname{Cornell University}, \orgaddress{\city{Ithaca}, \postcode{14853}, \state{New York}, \country{United States}}}

\affil[2]{\orgdiv{Department of Physics}, \orgname{University of Toronto}, \orgaddress{\city{Toronto}, \state{Ontario}, \country{Canada}}}

\affil*[3]{\orgname{Canadian Institute for Advanced Research}, \orgaddress{\city{Toronto},\state{Ontario}, \country{Canada}}}

\maketitle

%\linenumbers

\begin{natureabstract}

The thermal Hall effect has been observed in a wide variety of magnetic insulators \cite{hirschbergerLargeThermalHall2015,yokoiHalfintegerQuantizedAnomalous2021,czajkaOscillationsThermalConductivity2021,dokiSpinThermalHall2018,boulangerThermalHallConductivity2022,chenLargePhononThermal2022,liPhononThermalHall2020,liPhononThermalHall2023,vallipuramRoleMagneticIons2024}, yet its origins remains controversial. While some studies attribute the effect to intrinsic mechanisms \cite{qinBerryCurvaturePhonon2012,zhangPhononHallViscosity2021,yePhononHallViscosity2021,zhangTopologicalMagnonsThermal2021,flebusPhononHallViscosity2023}, such as heat carriers with Berry curvature, others propose extrinsic mechanisms \cite{chenEnhancedThermalHall2020,flebusChargedDefectsPhonon2022,guoResonantThermalHall2022}, such as heat carriers scattering off crystal defects. Even the nature of the heat carriers is unknown: magnons, phonons, and fractionalized spin excitations have all been proposed. Resolving these issues is essential for the study of quantum spin liquids, and particularly for \rcl, where a quantized thermal Hall effect has been attributed to Majorana edge modes \cite{kasaharaMajoranaQuantizationHalfinteger2018,bruinRobustnessThermalHall2022}. Here, we use ultrasonic measurements of the acoustic Faraday effect to demonstrate that the phonons in \rcl~have Hall viscosity---a non-dissipative viscosity that rotates phonon polarizations and deflects phonon heat currents. We show that phonon Hall viscosity produces an intrinsic thermal Hall effect that quantitatively accounts for a significant fraction of the measured thermal Hall effect in \rcl: the thermal Hall effect in \rcl is due to phonons \textit{and} it is intrinsic. More broadly, we demonstrate that the acoustic Faraday effect is a powerful tool for detecting phonon Hall viscosity and the associated phonon Berry curvature, offering a new way to uncover and study exotic states of matter that elude conventional experiments.

\end{natureabstract}

\section{Introduction}

Thermal transport has emerged as the dominant experimental technique in the search for charge-neutral quasiparticles~\cite{katsuraTheoryThermalHall2010,zhangThermalHallEffects2024}. The observation of a thermal Hall effect---where a temperature gradient builds up perpendicular to both a magnetic field and a heat current---has been used as evidence for fractionalized quasiparticles in quantum magnets such as Tb$_2$Ti$_2$O$_7$~\cite{hirschbergerLargeThermalHall2015}, \rcl~\cite{yokoiHalfintegerQuantizedAnomalous2021,czajkaOscillationsThermalConductivity2021}, and several kagome antiferromagnets~\cite{watanabeEmergenceNontrivialMagnetic2016,dokiSpinThermalHall2018}. Central to these claims is the assumption that phonons themselves do not generate a thermal Hall effect. This is a reasonable assumption, as phonons carry neither electric charge nor spin and thus should not interact directly with magnetic fields.

This assumption has recently been called into question due to the discovery of relatively large thermal Hall effects in conventional antiferromagnets that are unlikely to host spinons~\cite{boulangerThermalHallConductivity2022,chenLargePhononThermal2022}, as well as in non-magnetic insulators where phonons are the only heat carrier~\cite{liPhononThermalHall2020,liPhononThermalHall2023,vallipuramRoleMagneticIons2024}. This suggests that phonons may be responsible for the thermal Hall effect in at least a subset of these materials. 

How can phonons generate a thermal Hall effect? Intrinsic mechanisms invoke phonon Berry curvature~\cite{qinBerryCurvaturePhonon2012,zhangPhononHallViscosity2021,yePhononHallViscosity2021,flebusPhononHallViscosity2023}, whereas extrinsic mechanisms invoke phonon skew scattering~\cite{chenEnhancedThermalHall2020,flebusChargedDefectsPhonon2022,guoResonantThermalHall2022}. While a large number of theoretical proposals have been put forward~\cite{qinBerryCurvaturePhonon2012,zhangPhononHallViscosity2021,yePhononHallViscosity2021,flebusPhononHallViscosity2023,chenEnhancedThermalHall2020,flebusChargedDefectsPhonon2022,guoResonantThermalHall2022,zhangTopologicalMagnonsThermal2021,chernSignStructureThermal2021,gordonTheoryFieldrevealedKitaev2019}, the only experimental probe thus far has been thermal transport: applying heat currents and measuring the resultant temperature gradients. A key limitation of thermal transport is its inability to distinguish between intrinsic and extrinsic mechanisms, and this has made it difficult to distinguish between proposed origins of the thermal Hall effect at even a coarse level.

\begin{figure}[H]
	\begin{center}
		\includegraphics[width=.8\textwidth]{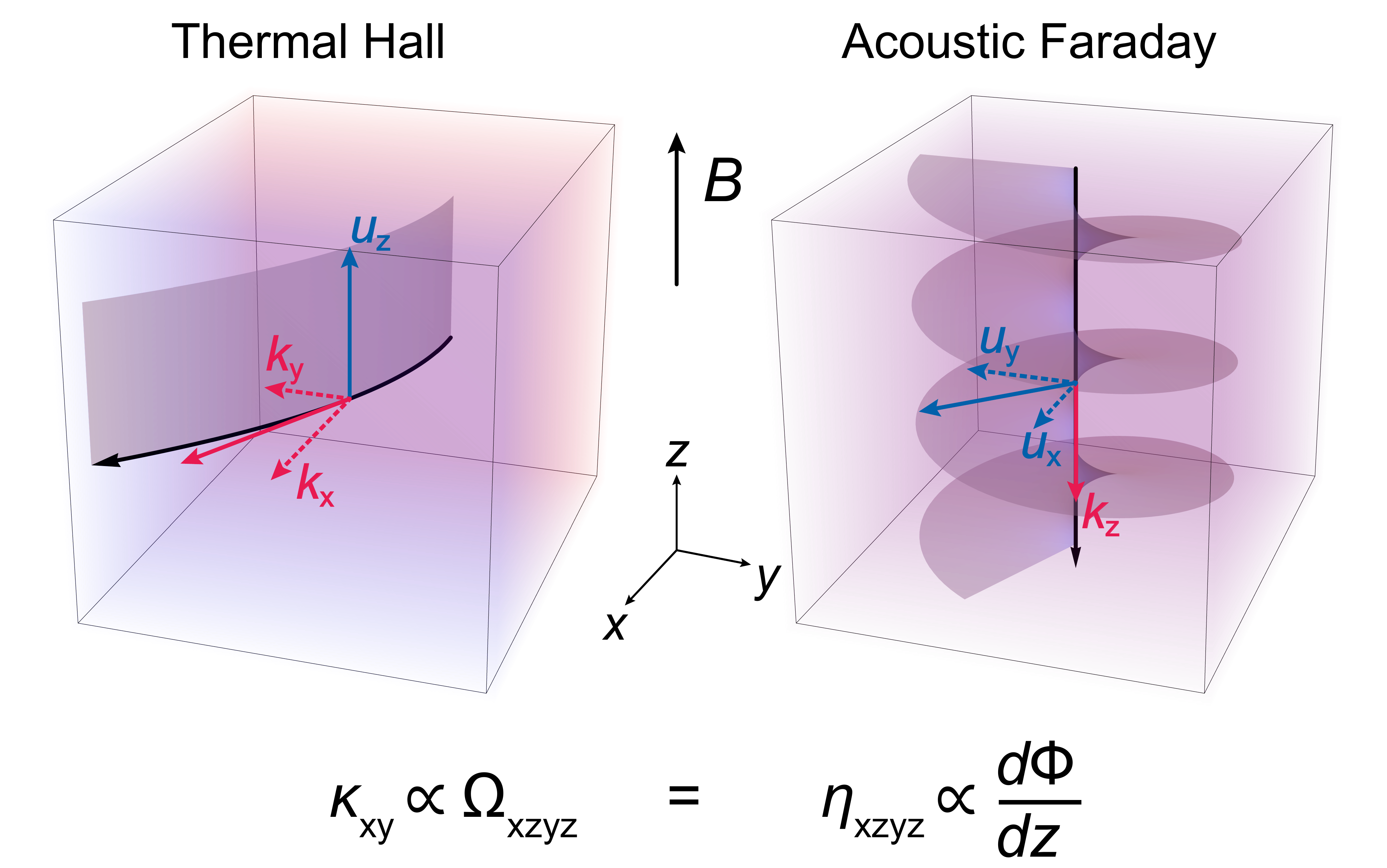}
	\end{center}
	\caption{\textbf{The relationship between the thermal Hall and acoustic Faraday effects.} Applying a magnetic field $B$ parallel to $z$ breaks time reversal symmetry and allows for both phonon Berry curvature, $\Omega_{xzyz}$, and phonon Hall viscosity, $\eta_{xzyz}$. Phonon Hall viscosity can be thought of as the long-wavelength manifestation of the microscopic Berry curvature---the equals sign indicates that these two quantities are equivalent~\cite{avronViscosityQuantumHall1995,barkeshliDissipationlessPhononHall2012}. These schematics illustrate how phonon Berry curvature and Hall viscosity affect the phonon wavevector, $\vec{k}$ (blue lines), and phonon polarization vector, $\vec{u}$ (magenta lines). On the left, a red-blue color gradient indicates an applied temperature gradient along $x$. Phonon Berry curvature deflects the phonon heat current, generating a heat current along $y$ (the black line indicates the phonon trajectory). On the right, the polarization of acoustic phonons traveling along $z$ rotates at a rate that is proportional to the Hall viscosity ($\Phi$ is the polarization rotation angle, and the black line is the phonon propagation vector).}
	\label{fig:faraday_thermal}
\end{figure}

Here, we take advantage of the fact that phonon Berry curvature not only causes an intrinsic thermal Hall effect, but also generates a dissipationless viscous force known as phonon Hall viscosity~\cite{avronViscosityQuantumHall1995,barkeshliDissipationlessPhononHall2012}. Phonon Hall viscosity rotates the polarization of transverse-polarized phonons as they propagate (see \autoref{fig:faraday_thermal}). This is called the acoustic Faraday effect, and it is analogous to how the optical Faraday effect rotates the polarization of light. The principal advantage of the acoustic Faraday effect over thermal transport is its reliance on coherent phonons. Because the measurement is sensitive only to a specific wavevector, phase, and frequency, it provides a direct probe of intrinsic polarization rotation, effectively filtering out the phase-incoherent scattering events that contribute to extrinsic transport.

We measure the acoustic Faraday effect in \rcl~and extract the phonon Hall viscosity. We find that the Hall viscosity is peaked near the magnetic phase boundary where the thermal Hall effect is maximized. The existence of phonon Hall viscosity in \rcl~suggests a non-zero phonon contribution to the thermal Hall effect, challenging interpretations of the thermal Hall effect solely in terms of Majorana fermions~\cite{kasaharaMajoranaQuantizationHalfinteger2018,yokoiHalfintegerQuantizedAnomalous2021,bruinRobustnessThermalHall2022} or other fractionalized spin excitations~\cite{czajkaOscillationsThermalConductivity2021}. We suggest that phonons acquire Hall viscosity by coupling to spins through the large spin-orbit coupling present in \rcl. This general mechanism is likely responsible for the unusual thermal Hall effects reported in many other magnetic insulators.

\section{Results}

\subsection{Measuring the acoustic Faraday effect}

%and its strength is characterized by a Verdet constant with dimensions of radians per tesla-meter. 

We first describe how phonon Hall viscosity alters how sound propagates through a solid. In crystals with high enough symmetry, and for high-symmetry propagation directions, the two transverse sound modes are degenerate, with the same speed of sound. This is the case for sound propagating along the $c$ axis of \rcl, which is the configuration in our experiment. The degeneracy of these two modes is lifted by Hall viscosity, resulting in left- and right-circularly polarized sound waves that propagate at different speeds. This gives rise to an acoustic Faraday effect---the rotation of linearly polarized transverse sound waves (see \autoref{fig:faraday_thermal}). This phenomenon is well documented in materials like yttrium iron garnet (YIG) \cite{kittelInteractionSpinWaves1958,guermeurRotationPlanPolarisation1968,wangAcousticFaradayRotation1971}, Cr$_2$O$_3$ \cite{boiteuxAcousticalFaradayEffect1971}, Tb$_3$Ga$_5$O$_{12}$ \cite{sytchevaAcousticFaradayEffect2010} (which also has a thermal Hall effect \cite{strohmPhenomenologicalEvidencePhonon2005}), CeAl$_2$ \cite{luthiMagnetoelasticityRotationalInvariance1979}, and in superfluid $^3$He-B \cite{leeDiscoveryAcousticFaraday1999}.

\begin{figure}[H]
	\begin{center}
			\includegraphics[width=.9\textwidth]{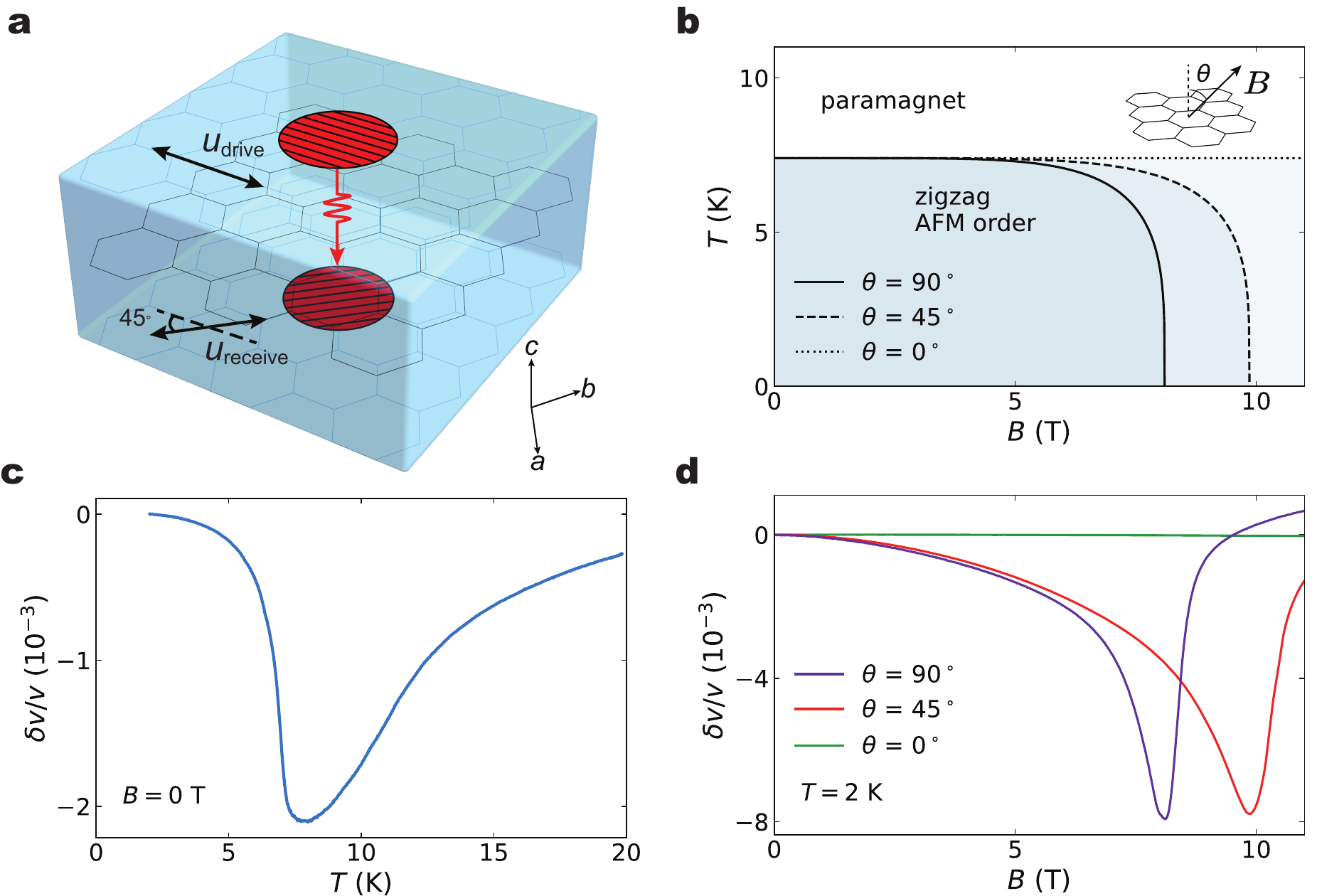}
	\end{center}
	\caption{\textbf{Experimental geometry and sample characterization.} \textbf{a}, A piezoelectric drive transducer (upper red circle) sends a pulse of traverse sound (red line) along the $c$ axis of a single-crystal \rcl~sample (indicated schematically by the honeycomb). The pulse is detected by a second, receive transducer (lower red circle) with its polarization rotated 45$^{\circ}$ relative to the drive transducer. \textbf{b}, Schematic temperature-field phase diagram of \rcl~for different field orientations. The inset defines $\theta$ as the angle between the magnetic field $\mathbf B$ and the $c$ axis (rotation is in the $bc$ plane). We show the phase boundary for field along the $b$ axis ($\theta = 90^{\circ}$, sold line), field along the $c$ axis ($\theta = 0^{\circ}$, short dashed line), and field $45^{\circ}$ between the $b$ and $c$ axes (long dashed line). \textbf{c}, The relative change in sound velocity as a function of temperature across $T_{\rm{N}} = 7.5$ K. The sample exhibits a single magnetic phase transition, with no features near 14 K that would indicate a secondary structural and magnetic phase~\cite{caoLowtemperatureCrystalMagnetic2016,kimStructuralTransitionMagnetic2024}. \textbf{d}, The relative change in sound velocity as a function of magnetic field across the critical field:  $B_c = 8$ T for $B||b$, and $B_c = 10$ T for $\theta = 45^{\circ}$. For $B||c$, the sample remains in the ordered phase up to 12 tesla. The single phase transition indicates that the field is well-aligned with the $b$ axis~\cite{balzFieldinducedIntermediateOrdered2021}.}
	\label{fig:geometry_coupling}
\end{figure}

The acoustic Faraday effect is measured using shear-polarized piezoelectric transducers to generate pulses of linearly polarized, transverse sound (see \autoref{fig:geometry_coupling}a). These sound waves travel the length of the sample and are detected by a second shear-polarized piezoelectric transducer on the opposite face from the drive transducer. The amplitude of sound is maximum when the sound-wave polarization and receive-transducer polarization are parallel, and minimum when they are perpendicular. In general, varying the external magnetic field changes the Hall viscosity, which modifies the amount of Faraday rotation and, consequently, the amplitude of sound detected by the receive transducer.

One complication is that the amplitude of sound detected by the receive transducer depends not only on the amount of Faraday rotation, but also on sound attenuation. The attenuation can vary significantly with magnetic field, especially near a magnetic phase transition. This is particularly relevant for \rcl, where a spin-liquid state has been proposed near the critical field where long-range magnetic order is suppressed ($\approx$8 tesla for in-plane magnetic fields, see \autoref{fig:geometry_coupling}b). Importantly for our experiment, sound attenuation is an \textit{even} function of magnetic field---it depends on only the magnitude of the magnetic field, and not on the sign. In contrast, the acoustic Faraday effect is an \textit{odd} function of magnetic field, changing sign with field reversal. To isolate the Faraday contribution, we measure the received sound amplitude for both positive and negative magnetic fields and then antisymmetrize the data. 

Single-crystal samples of \rcl were prepared by sputtering piezoelectric ZnO transducers on two opposite $c$-axis faces. One transducer was excited using $50$ ns bursts of $\approx 1.5$ GHz radiofrequency voltage, and the signal received at the second transducer was amplified and digitized using an oscilloscope. The amplitude and phase of the received signal were extracted using a software lockin. Further experimental details can be found in the Methods. The high quality of our samples, and their rhombohedral crystal structure at low temperature \cite{kimStructuralTransitionMagnetic2024}, is demonstrated by their single phase transition near 7.5 kelvin in zero magnetic field (\autoref{fig:geometry_coupling}c).

\subsection{The acoustic Faraday effect of \rcl~for $B||c$}

\begin{figure}[H]
	\begin{center}
		\includegraphics[width=1\textwidth]{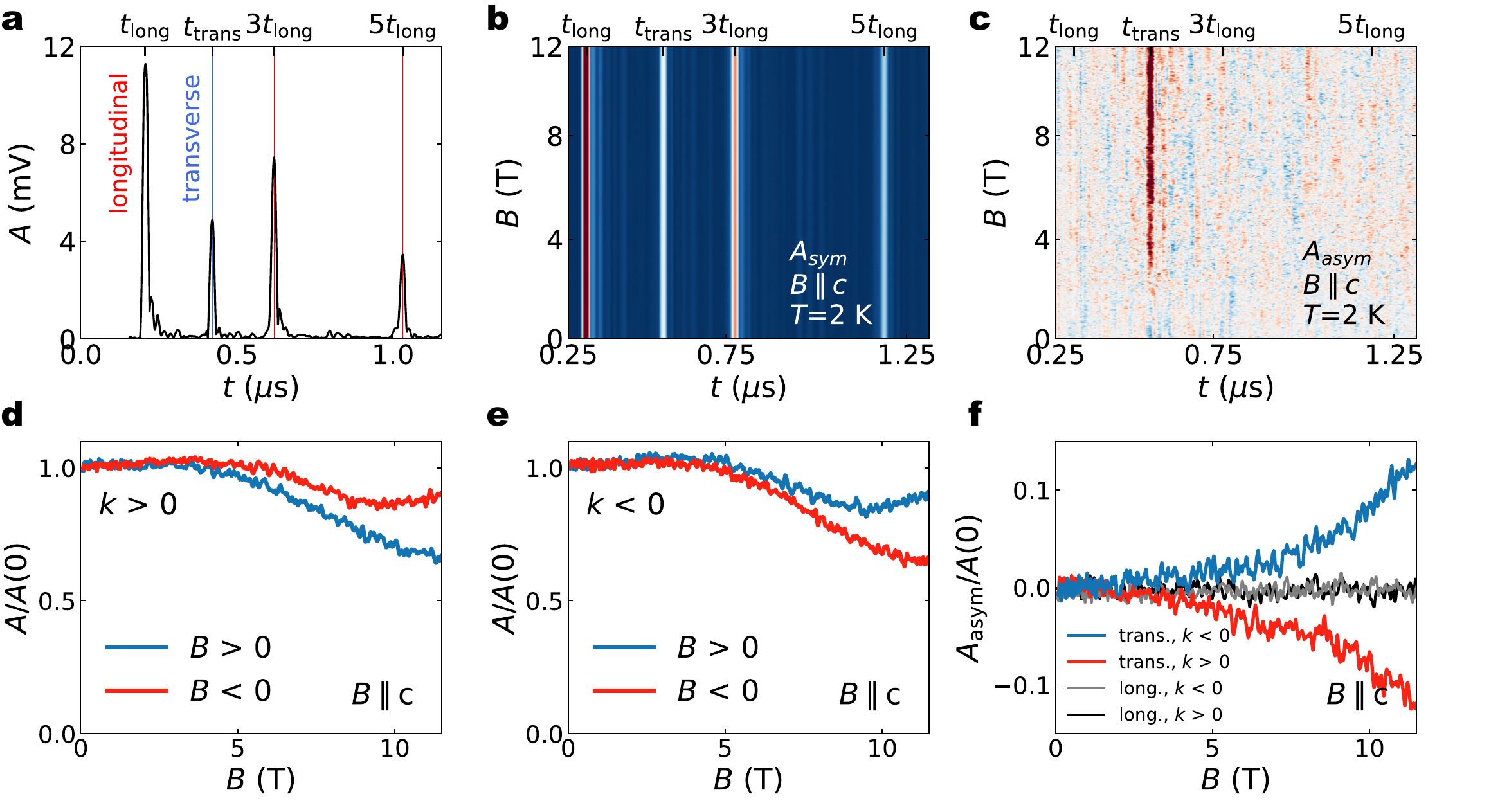}
	\end{center}
	\caption{\textbf{Isolating the acoustic Faraday effect.} \textbf{a}, The raw voltage amplitude $A$ detected using the setup shown in \autoref{fig:geometry_coupling}a, for $B||c = 10$ T and $T = 2$ K. Longitudinal and transverse pulses are identified based on their known speeds of sound~\cite{hauspurgFractionalizedExcitationsProbed2024, lebertAcousticPhononDispersion2022}. The first longitudinal pulse arrives at $t_{\rm long}$, with echoes of this signal arriving at $3t_{\rm long}$ and $5t_{\rm long}$. The transverse pulse arrives at $t_{\rm trans}$ and is clearly separated in the time domain from the longitudinal signal. \textbf{b}, The field-symmetrized data show all longitudinal and transverse signals from 0 to 12 tesla. \textbf{c}, The field-antisymmetrized data contains only transverse signal because the Faraday effect rotates the transverse sound polarization in opposite directions for $\pm \mathbf B$, whereas it cannot rotate longitudinal polarization. \textbf{d}, The amplitude of transverse sound, normalized to its zero-field value, as a function of magnetic field for ${\mathbf B}||c$ at $T = 2$ K. The magnetic field suppresses the amplitude more for positive field than for negative field. \textbf{e}, This behaviour switches when the propagation direction is switched, as expected for a Faraday effect. \textbf{f}, The field-antisymmetrized amplitude for both propagation directions. The transverse signal shows the characteristics of a Faraday effect, going to zero at ${\mathbf B} = 0$, and switching sign when ${\mathbf k}\rightarrow-{\mathbf k}$. The longitudinal signal, in contrast, is zero at all fields for both propagation directions.}
	\label{fig:faraday}
\end{figure}

We first show data taken with the magnetic field swept from 0 to 12 tesla along the $c$ axis, at a fixed temperature of 2 kelvin. In this configuration, the critical field to suppress magnetic order is over 30 tesla~\cite{modicScaleinvariantMagneticAnisotropy2021,zhouPossibleIntermediateQuantum2023}, and the entire measurement is deep within the ordered antiferromagnetic state. This allows us to first demonstrate the acoustic Faraday effect in \rcl~without the additional complication of the magnetic phase transition (this configuration also has a thermal Hall effect, see Le Francois \textit{et al.}~\cite{lefrancoisEvidencePhononHall2022}). \autoref{fig:faraday}a shows the amplitude of transmitted sound as a function of time at a fixed field of $B = 10$ T along the $c$-axis and in the magnetically ordered state at $T = 2$ K. The mixed-mode transducers generate and detect both longitudinal and transverse sound waves, which travel at different speeds. The longitudinal and transverse modes are visible as distinct pulses in the time-series data, and we use both to validate our experimental method. \autoref{fig:faraday}b (c) shows data that have been symmetrized (antisymmetrized) in magnetic field. Antisymmetrization completely removes the longitudinal signal, whereas transverse signal remains. This is consistent with the transverse signal arising due to the acoustic Faraday effect.

\autoref{fig:faraday}d shows the amplitude of the transverse signal (at $t = t_{\rm trans.}$) as a function of magnetic field for both positive and negative field directions ($\pm \mathbf{B}$), and panel e shows the same but for the opposite sound propagation direction. The asymmetry between $+\mathbf B$ and $- \mathbf B$ switches sign when the sound propagation direction $\mathbf{k}$ is switched to $-\mathbf{k}$ (by swapping the drive and receive cables at the top of the cryostat). The change in sign of the antisymmetric signal when changing $\mathbf{k}$ to $-\mathbf{k}$, the fact that the antisymmetric signal is non-zero only for transverse sound, and the vanishing of the antisymmetric signal at zero magnetic field, all confirm that the antisymmetric signal is caused by the acoustic Faraday effect. Converting our antisymmetric signal into a polarization rotation angle, we find that the polarization rotates by 20$^{\circ}$/mm at 2 K and at 12 T with $B||c$ (see Extended Data).

Having demonstrated that our procedure isolates the acoustic Faraday effect, we characterized its temperature, field, and field-angle dependence. These results are corroborated by several consistency checks and by measurements on additional samples (see Methods).

\subsection{Extracting the Hall viscosity of \rcl}

We now extract the phonon Hall viscosity by modeling our acoustic Faraday data using the elastic wave equation with viscous contributions included. The relationship between stress, $\bm{\sigma}$, and strain, $\bm{\varepsilon}$, in a solid---Hooke's law---is modified at finite frequency by introducing the viscosity tensor $\hat{\eta}$: 
\begin{equation}
	\sigma_{ij} = c_{ijkl}\varepsilon_{kl}+\eta_{ijkl}\dot{\varepsilon}_{kl},
	\label{eq:hooke}
\end{equation}
where $\hat{c}$ is the elastic tensor and $\bm{\dot{\varepsilon}}$ is the time derivative of strain. While the most familiar viscosity components are dissipative, time-reversal symmetry breaking (either intrinsic or from an applied magnetic field) allows for additional, \textit{non}-dissipative ``Hall'' viscosities. A magnetic field component along the $c$ axis of \rcl~activates the $\eta_{xzyz}$ component. 

When incorporated into the elastic wave equation, $\eta_{xzyz}$ couples transverse waves polarized along $y$ to transverse waves polarized along $x$ for waves propagating along the $z$ direction. This coupling rotates linearly polarized transverse waves and produces an acoustic Faraday effect. Note that $\eta_{xzyz}$ is allowed as long as there is \textit{any} component of field along $c$. Other Hall viscosities are allowed when other field components are present---see Methods for a full symmetry analysis of the allowed Hall viscosities. The wave equation with Hall viscosity included is 
\begin{equation}
	\rho \omega^2 \vec{u} = k^2 \begin{bmatrix}
		c_{xzxz} & i\omega \eta_{xzyz} \\
		-i\omega \eta_{xzyz} & c_{yzyz}
	\end{bmatrix}	 \vec{u},
	\label{eq:wave}
\end{equation}
where $\rho$ is the material density, $\omega$ is the sound frequency, $\vec{u}$ is the sound polarization vector, $k$ is the sound wavenumber, and $c_{xzxz}\equiv c_{55}$ and $c_{yzyz}\equiv c_{44}$ are elastic constants. 

\begin{figure}[H]
	\begin{center}
			\includegraphics[width=1\textwidth]{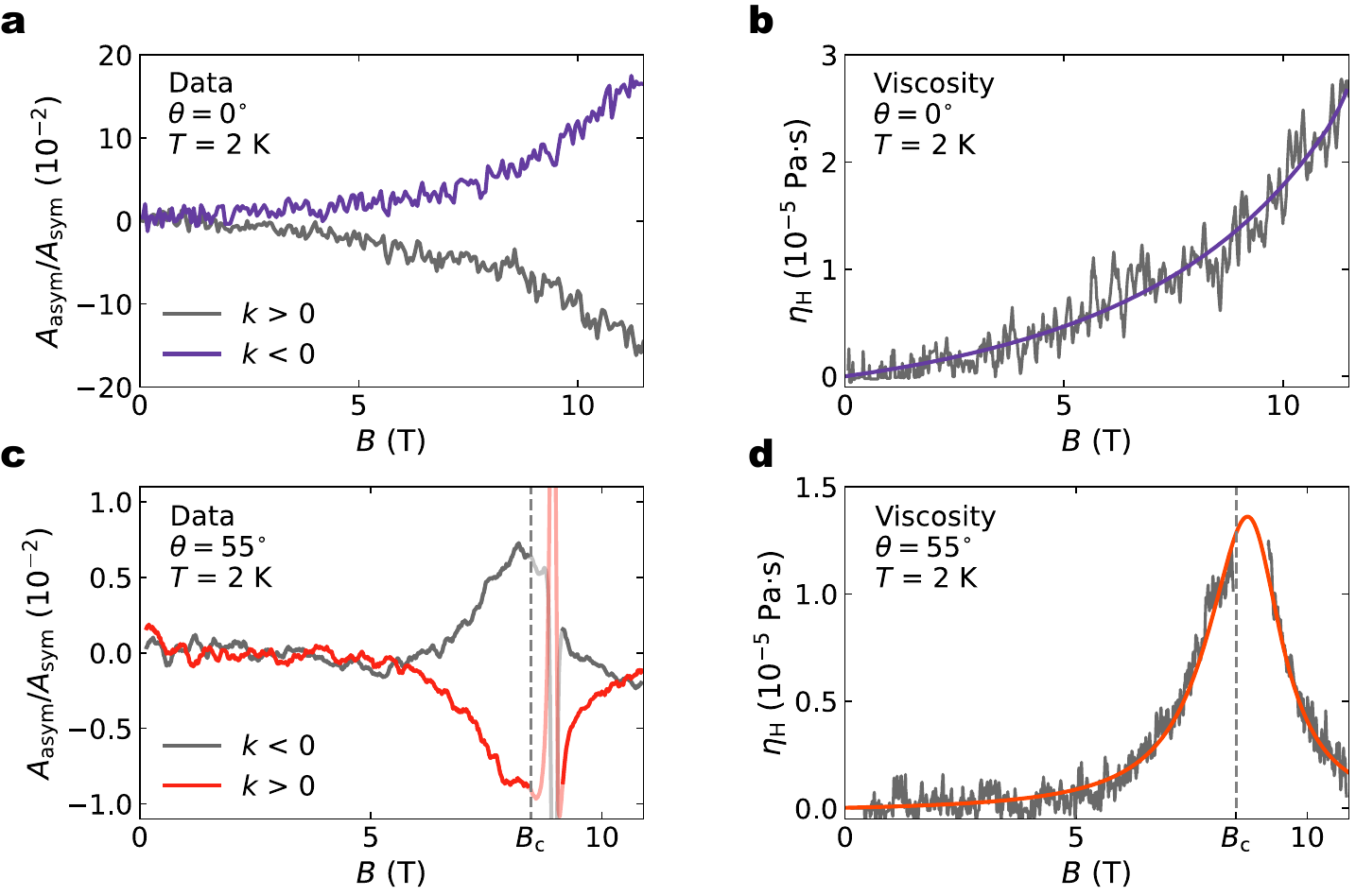}
	\end{center}
	\caption{\textbf{The phonon Hall viscosity of \rcl.} \textbf{a}, The field-antisymmetrized Faraday signal normalized to the field-symmetrized signal, as a function of magnetic field for $\mathbf{B}||c$, at $T = 2$ K, for $\pm \mathbf{k}$. \textbf{b}, The Hall viscosity extracted from the antisymmetric signal using \autoref{eq:wave}. The solid line is a guide to the eye. Panels \textbf{c} and \textbf{d} show the same data and analysis, but for magnetic field rotated $55^{\circ}$ toward the $b$ axis. The sharp feature slightly above $B_c$ in panel \textbf{c} is likely a result of the small absolute signal size and the rapid change in speed of sound near this field. We have truncated this feature from the viscosity in panel \textbf{d}.}
	\label{fig:model}
\end{figure}

\autoref{fig:model}a and c show the antisymmetrized Faraday rotation data taken with $B||c$ and $B$ rotated $55^{\circ}$ toward the $b$ axis, respectively. Both data sets are taken at 2 kelvin: the $B||c$ data is entirely in the magnetically ordered state, whereas the $\theta = 55^{\circ}$ data crosses the antiferromagnetic-to-paramagnetic transition at $B_{\rm c} = 8.5$ T.

For $B||c$, $c_{44}$ and $c_{55}$ are equal, and we measure this elastic constant independently using the time-of-flight between echoes (see \autoref{fig:faraday}a). The Hall viscosity $\eta_{xzyz}$ is the only unknown in \autoref{eq:wave} and is therefore directly determined by the antisymmetric amplitude (recall that field-\textit{symmetric} changes in amplitude are removed through the antisymmetrization procedure). 

Tilting the magnetic field away from the $c$ axis breaks the three-fold rotational symmetry of the lattice, inducing acoustic birefringence ( $c_{44}\neq c_{55}$). This produces elliptical polarization, which means that the effects of birefringence cannot be entirely removed via antisymmetrization. However, birefringence is intrinsically symmetric in field; the emergence of a field-antisymmetric signal can only be produced by the antisymmetric Hall viscosity. Thus the presence of birefringence produces only a small quantitative shift in our extracted Hall viscosity (see Methods).

We show the phonon Hall viscosity extracted from the antisymmetric data using \autoref{eq:wave} in \autoref{fig:model}b and d. For $B||c$, the Hall viscosity increases continuously as a function of magnetic field up to $\eta_{xzyz}=0.03$ mPa$\cdot$s at 12 T. For $\theta = 55^{\circ}$, the Hall viscosity peaks just above the critical field of 8.5 T, reaching a value of approximately 0.013 mPa$\cdot$s. 

\subsection{Comparison to the thermal Hall effect}

How do our measurements of phonon Hall viscosity relate to the thermal Hall effect in \rcl? The Hall viscosity can be interpreted in two ways. In an acoustic Faraday experiment with magnetic field along $z$, $\eta_{zxzy}$ mixes a sound wave propagating along $z$ and polarized along $x$ with a sound wave propagating along $z$ and polarized along $y$. In a thermal transport experiment with magnetic field along $z$, $\eta_{xzyz}$ takes phonon heat flow along $x$ and deflects it to heat flow along $y$. By symmetry, the viscosity elements $\eta_{zxzy}$ and $\eta_{xzyz}$ are equal and thus our observation of phonon Hall viscosity requires a phonon contribution to the thermal Hall effect.

We can estimate this contribution using our measurement of $\eta_{xzyz}$ and the equations for acoustic energy transport. This is a classical, kinetic model for thermal conductivity---a detailed derivation is given in the Methods. Intuitively, heat is carried by a thermal population of phonons, whereas the acoustic Faraday measurement probes the Hall viscosity of a single, low-frequency phonon mode. Assuming that the Hall viscosity is frequency independent, we can relate the thermal Hall conductivity to the phonon Hall viscosity through the specific heat: $\kappa_{xy} = \frac{\eta_{xzyz}}{\rho} C$, where $\rho$ and $C$ are \rcl's density and specific heat, respectively. Note that this phonon Hall conductivity is intrinsic---it is independent of the phonon mean free path. This Hall conductivity can be compared with the longitudinal phonon thermal conductivity: $\kappa_{xx} = \overline{v}_s l C$, where $\overline{v}_s$ is the average speed of sound and $l$ is the phonon mean free path. The ratio of phonon Hall to phonon longitudinal thermal conductivities is then
\begin{equation}
	\frac{\kappa_{xy}}{\kappa_{xx}} = \frac{\eta_{xzyz}}{\rho v_s l} = \frac{C}{ \kappa_{xx}}\frac{\eta_{xzyz}}{\rho}.
	\label{eq:ratio}
\end{equation}

Using our measured value of $\eta_{xzyz}=3\times10^{-5}~$Pa$\cdot$s, and the measured values of $\kappa_{xx}$ and the specific heat at $10$ K with $\mathbf{B}||c$~\cite{lefrancoisEvidencePhononHall2022,tanakaThermodynamicEvidenceFieldangledependent2022}, we estimate $\kappa_{xy}/\kappa_{xx} = 2.5\times 10^{-4}$---within a factor of 6 of what is observed in thermal transport experiments under the same conditions by LeFrancois \textit{et al.}~\cite{lefrancoisEvidencePhononHall2022} (as discussed below, the total contribution from all phonon branches and viscosity tensor elements resolves this difference). Our analysis also determines the absolute sign of the Hall viscosity to be positive ($\eta_{xzyz} > 0$), which is consistent with the positive thermal Hall effect (see Methods). 

This estimate represents a conservative lower bound on the intrinsic phonon contribution to $\kappa_{xy}$, as it accounts for only 1 of the 3 acoustic branches and a single viscosity tensor element. Since all 4 symmetry-allowed viscosities are predicted to be of comparable magnitude \cite{dhakalTheoryIntrinsicPhonon2025}, the cumulative contribution from all 3 branches is sufficient to account for the total measured thermal Hall effect. This quantitative agreement is consistent with recent first-principles calculations \cite{dhakalTheoryIntrinsicPhonon2025}. Furthermore, by performing acoustic Faraday measurements from 1 to 1.7 GHz, we find that the Hall viscosity grows with increasing frequency (see Extended Data). This suggests that the Hall viscosity measured in our experiment at $\approx 1$ GHz may serve as a lower bound for the Hall viscosity that contributes to thermal transport, which probes a broader energy distribution of phonons.

\begin{figure}[H]
	\begin{center}
		\includegraphics[width=\textwidth]{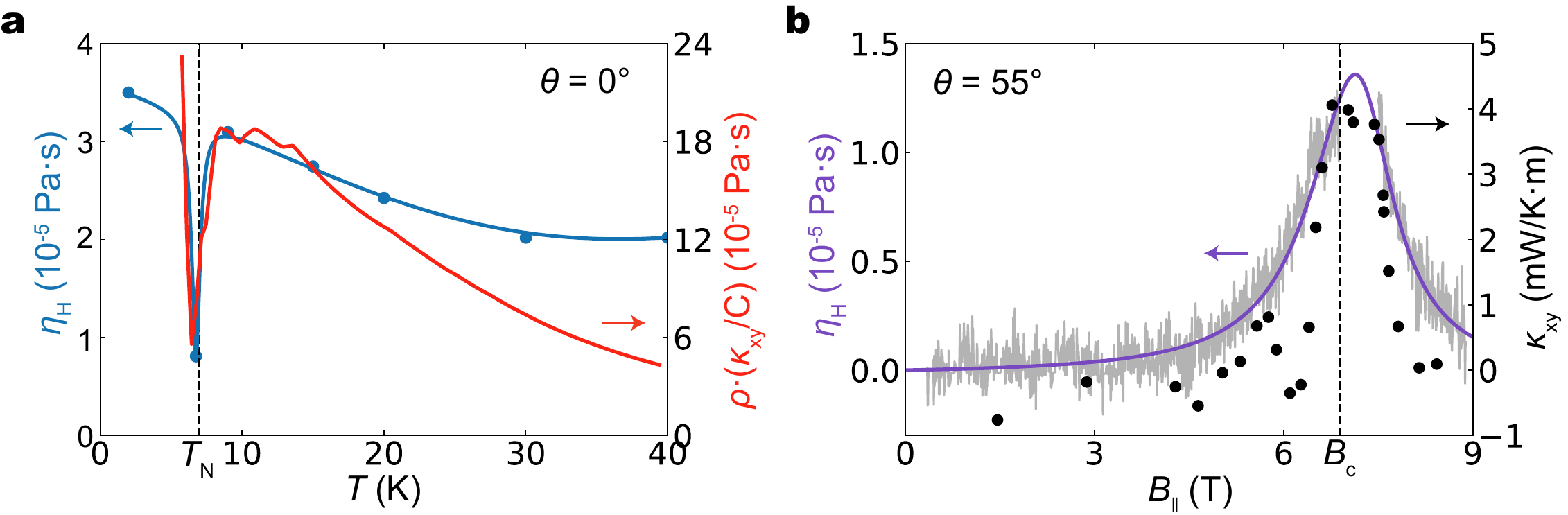}
	\end{center}
	\caption{\textbf{Temperature and magnetic field dependence of the phonon Hall viscosity and thermal Hall effect.} \textbf{a}, The phonon Hall viscosity as a function of temperature for $\mathbf{B}||c = 12$ T. Blue circles are measured data points, and the blue line is a guide to the eye. The quantity $\rho \kappa_{xy}/C$---calculated using $\kappa_{xy}$ from LeFrancois \textit{et al.}~\cite{lefrancoisEvidencePhononHall2022} and $C$ from Widman \textit{et al.}~\cite{widmannThermodynamicEvidenceFractionalized2019}---is plotted in red for comparison. \textbf{b}, The phonon Hall viscosity measured with $\mathbf{B}$ rotated 55$^{\circ}$ from $c$ toward $b$, plotted as a function of in-plane magnetic field at $\it{T}=$ 2 K. The $\kappa_{xy}$ data are measured at $\it{T}=$ 4.9 K with $\mathbf{B}$ rotated 45$^\circ$ away from $c$, and are taken from Kasahara \textit{et al.}~\cite{kasaharaMajoranaQuantizationHalfinteger2018}.}
	\label{fig:temp}
\end{figure}

Next, we analyze the temperature dependence of the Hall viscosity for $B||c$. \autoref{fig:temp}a shows that, aside from a sharp drop near the magnetic phase transition at 7.5 K, the Hall viscosity is only weakly temperature dependent up to 40 kelvin. This clearly demonstrates that phonon Hall viscosity in \rcl~is not due to the coherent hybridization of acoustic phonons and magnons into chiral ``magnetopolarons''~\cite{kittelInteractionSpinWaves1958}---the conventional mechanism of the acoustic Faraday effect. We can compare the temperature dependence of the Hall viscosity to the thermal Hall effect by plotting $\eta_{xzyz}$ alongside $\rho \kappa_{xy}/C$, which has units of viscosity. These quantities are the same order of magnitude and have qualitatively similar temperature dependencies: both showing minima near $T_{\rm N}$ and extend to temperatures much greater than $T_{\rm N}$. The drop in $\eta_{xzyz}$ near $T_{\rm{N}}$ may be intrinsic, or it may be due to strong phonon scattering due to thermal fluctuations of the order parameter at the antiferromagnetic phase transition (although $\eta_{xzyz}$ itself does not depend on the phonon mean free path, our ability to measure it does when the mean free path of our ultrasonic phonons becomes much shorter than the sample size, which happens near $T_{\rm{N}}$).

Finally, in \autoref{fig:temp}b we compare the magnetic field dependence of the Hall viscosity at $\theta = 55^{\circ}$ to the field dependence of the thermal Hall effect measured at a similar angle. Both quantities grow above 5 tesla and are peaked near the critical field where long-range order is suppressed. While a quantitative comparison between Hall viscosity and thermal Hall effect is not possible for this field orientation (thermal Hall is sensitive to 9 viscosity tensor components at this angle, whereas acoustic Faraday measures only 1), \autoref{fig:temp}b clearly shows that both quantities are sensitive to the proximity of the critical field where order is suppressed. It is worth noting that, while crystal symmetry constrains $\kappa_{xy}$ to be zero when the magnetic field purely along $b$, the same symmetry argument does not require all viscosity components to vanish for this field orientation (this is discussed further in Dhakal \textit{et al.}~\cite{dhakalTheoryIntrinsicPhonon2025}).

\section{Discussion}

We have demonstrated that phonons can account for a significant fraction---if not all---of the observed thermal Hall effect in \rcl when a component of the magnetic field is perpendicular to the planes ($B_z$). The relevance of these results extends directly to measurements of the ``planar'' thermal Hall effect, where the magnetic field is applied purely in plane \cite{yokoiHalfintegerQuantizedAnomalous2021,czajkaPlanarThermalHall2023,imamuraMajoranafermionOriginPlanar2024}. While the acoustic Faraday effect requires a $B_z$ component to probe $\eta_{xzyz}$, the symmetry of \rcl allows for a common microscopic origin for both planar and conventional thermal Hall effects \cite{kurumajiSymmetrybasedRequirementMeasurement2023}. Therefore, if phonon Hall viscosity accounts for the out-of-plane effect, the same mechanism can naturally account for the effect induced by an in-plane field. This connection is supported by theoretical calculations indicating that the thermal Hall effect induced by spin-lattice coupling is comparable for both in-plane and out-of-plane field directions \cite{liThermalHallEffect2022}. Furthermore, we observe that, for tilted magnetic fields, the viscosity is maximized near the critical field ($B_c$) where magnetic order is suppressed. This is the same field where the thermal Hall effect reaches its maximum in both tilted and planar configurations, demonstrating that they share a common microscopic origin. 

Next, we turn to the question of \textit{why} the phonons in \rcl~have Hall viscosity. Our observation of phonon Hall viscosity at high temperatures suggests that long-range magnetic order is not required. Instead, Hall viscosity emerges from the strong coupling between short-range magnetic fluctuations and the lattice. At a microscopic level, sound waves distort the bond distances and angles of the crystal lattice, thereby modulating the magnetic exchange interactions between neighboring ruthenium ions \cite{kimCrystalStructureMagnetism2016,kaibMagnetoelasticCouplingEffects2021}. The resulting dynamics of the local moments feed back into the phonon system, generating Hall viscosity. 

This mechanism has been shown to be particularly effective in \rcl because strong spin-orbit coupling makes magnetic exchange highly sensitive to strain \cite{dhakalTheoryIntrinsicPhonon2025}. Because this mechanism relies on short-range magnetic correlations rather than long-range order \cite{dhakalTheoryIntrinsicPhonon2025}, it explains why a sizeable Hall viscosity---and thus thermal Hall effect \cite{lefrancoisEvidencePhononHall2022}---persists into the paramagnetic phase at high temperatures (this can also be thought of as strong coupling between phonons and paramagnons). It also explains why the Hall viscosity is maximized near $B_c$ with an in-plane field: both static spin susceptibility measurements and inelastic neutron scattering show that the magnetic state becomes ``soft'' at $B_c$~\cite{doShortrangeQuasistaticOrder2018,balzFiniteFieldRegime2019}, bringing magnetic excitations to low energies where they hybridize with the acoustic phonons that dominate thermal transport at low temperature. 

The measured phonon Hall viscosity compares well quantitatively with first-principles calculations of the spin-lattice coupling in \rcl by Dhakal \textit{et al.} \cite{dhakalTheoryIntrinsicPhonon2025}. Although they use a two-dimensional model that does not access the same viscosity component that we do in our experiment ($\eta_{xzyz}$), they find that all five in-plane viscosity components are the same order of magnitude---and, strikingly, the same order of magnitude as the value we measure for $\eta_{xzyz}$. 

Our measurement provides experimental evidence that the phonon Hall viscosity in \rcl~is indeed large enough to account for the thermal Hall effect, calling into question the existence of a quantized contribution from Majorana edge modes. Phonon Hall viscosity also likely explains the unusual thermal Hall effects found in many other magnetic insulators. However, non-magnetic materials, such as SrTiO$_3$, likely require a different mechanism.

Beyond clarifying the origin of the thermal Hall effect in \rcl, our measurements highlight the connection between phonon Hall viscosity and phonon Berry curvature---two concepts that have, until now, remained largely theoretical~\cite{avronViscosityQuantumHall1995,barkeshliDissipationlessPhononHall2012,yePhononHallViscosity2021, zhangPhononHallViscosity2021,flebusPhononHallViscosity2023}. Conventionally, viscosity is thought of as a long-wavelength property of sound that produces stress when a system is subject to time-varying strain (\autoref{eq:hooke} from above):
\begin{equation*}
	\sigma_{ij} = c_{ijkl}\varepsilon_{kl}+\eta_{ijkl}\dot{\varepsilon}_{kl}.
	\label{eq:hooke2}
\end{equation*}
This expression is remarkably similar to the stress-train relationship derived for a fully quantum theory of phonons \cite{avronViscosityQuantumHall1995,barkeshliDissipationlessPhononHall2012}:
\begin{equation}
	\sigma_{ij} = c_{ijkl} \varepsilon_{kl} + \Omega_{ijkl}\dot{\varepsilon}_{kl},
	\label{eq:feynman}
\end{equation}
where $\Omega_{ijkl}$ is the phonon Berry curvature. From these expressions, it is clear that phonon Hall viscosity and phonon Berry curvature are one and the same---phonon Hall viscosity is the long-wavelength manifestation of the microscopic phonon Berry curvature. It can be insightful to interpret  $\Omega_{ijkl}$ as a ``real-space'' Berry curvature---it gives rise to a geometric phase when the crystal lattice is moved around a closed path by an oscillating strain field, such as the one created by a sound wave. This is analogous to the ``momentum-space'' Berry curvature of Bloch electrons, which gain geometric phases when traversing closed paths in momentum space. Where Bloch electrons experience a non-dissipative force from Berry curvature that leads to the anomalous Hall effect, the lattice (i.e. phonons and sound waves) experiences a non-dissipative Hall viscosity that leads to the thermal Hall and acoustic Faraday effects. 

In general, a variety of mechanisms can generate phonon Berry curvature and Hall viscosity. For example, electron-phonon coupling can generate phonon Berry curvature and Hall viscosity in quantum Hall systems, topological insulators, and topological superconductors~\cite{barkeshliDissipationlessPhononHall2012}. Similarly for spin-lattice coupling in chiral spin liquids~\cite{zhangPhononHallViscosity2021}. Thus, our measurements do not rule out the possibility of spinon excitations in \rcl. They do, however, require that phonons account for at least a significant fraction of the thermal Hall effect. More generally, we have demonstrated that the acoustic Faraday effect is a direct probe of phonon Hall viscosity or, equivalently, phonon Berry curvature. Because phonon Hall viscosity is a property of many exotic phases, including topological superconductors and 3D quantum Hall states, acoustic Faraday rotation measurements will be a valuable tool for discovering and characterizing these states of matter.

%\bibliographystyle{sn-nature}
%\bibliography{literature}

\newpage

\section{Methods}

\subsection{Crystal growth and selection}
Single crystals of \rcl were grown using the method described in Kim \textit{et al.} \cite{kimRuCl3OtherKitaev2022}.

We first sort crystals by eye for untwinned specimens. Once identified, we further select for crystals without visible stacking faults by examining the crystal morphology. Sections that appear to have faults are removed by cleaving the sample. It is also necessary to ensure that there is enough overlapping area between the top and bottom surfaces so that there is a direct path along the $c$ axis for sound to travel between two transducers. Finally, samples are always handled using brushes---never tweezers---to ensure minimal structural degradation.   

To confirm our results we measure three samples of varying thickness, all grown by the same method. A representative piece of \rcl used for the ultrasound experiment is shown in Extended Data Fig. 1a. The naming convention is summarized in Extended Data Table 1. 

\subsection{Transducer fabrication}
We perform all depositions in a 4-gun Angstrom Engineering NexDep magnetron sputtering system, with the sample mounted above the sputtering targets.

We sputter thin-film ZnO piezoelectric transducers on cleaved (001) planes of single-domain \rcl crystals. The transducer consist of three layers: a bottom Ti/Pt layer (8nm/100nm), an $\sim 1~\mu$m layer of piezoelectric ZnO, and a top layer of Ti/Pt (8nm/100nm). The bottom metallic layer ensures adhesion of the ZnO to the sample surface and functions as a ground electrode for the transducer. The top metallic stack serves as an electrode for excitation and read out. Both metal layers are DC sputtered in a pure argon environment.

We deposit the piezoelectric ZnO layer by RF sputtering with a 2 inch diameter, 99.99\% purity ZnO target. We use a 1:3 ratio of oxygen to argon flow rates at a total pressure of 3 mTorr. The deposition rate is approximately 0.2 \AA/s using a power density of 21 W/in$^2$. 

The polarization of the transducer is determined by the $c$ axis orientation of the ZnO layer. For shear-polarized transducers, we use ``glancing angle deposition" \citemethods{fuGlancingAngleDeposition} to orient the ZnO $c$ axis away from sample surface normal. To achieve this, we position the sample as far from the sputtering target as our chamber allows, and keep the sample stage fixed (i.e. no rotation) during deposition. The angle of incidence of ZnO ions on the sample surface in our sputtering chamber is approximately 70$^\circ$ away from the surface normal. The resulting ZnO films have their $c$ axis tilted away from the sample surface normal along the straight line connecting the sample and target. This process results in a transducer that generates both compressional and shear stress in the sample. The polarization of the shear stresses is in the direction of the ZnO $c$ axis tilt. 

To facilitate antisymmetrization of the data as a function of magnetic field, the polarization of one transducer is rotated 45$^\circ$ with respect to the polarization of the other transducer: this allows differentiation between ``rotation left'' vs ``rotation right''. A precise 45$^\circ$ misalignment is not necessary to perform the antisymmetrization, but maximizes the size of antisymmetric signal in the presence of viscosity.

\subsection{Sample mounting}

We mount the \rcl samples on custom-designed, two-port PCBs using GE varnish. The attachment point is far from the ultrasonic transducer to minimize unwanted strain effects in the measurement. To minimize possible motion of the sample in an applied magnetic field, we further support the sample with pieces of microscope slide. We make electrical contact to the transducer bottom electrode with silver paint, connecting the electrode directly to the PCB ground plane. We wire the top electrode of the transducer to a coplanar waveguide using 25 $\mu$m diameter 99\% purity silver wire. Waveguides to both transducers terminate in MMCX connectors. \autoref{extfig:sample-photo}b and c show samples S1 and S2 after transducer deposition and mounting on the PCB for measurement.  

\section{Ultrasound Measurements}

\subsection{Fridge/Magnet}
Measurements are performed in an Oxford Instruments Teslatron system, with a 12 T superconducting magnet and variable temperature insert. The sample is loaded onto an Oxford Heliox He$^3$ probe equipped with custom low-loss cupronickle coaxial cables. 

\subsection{Measurement Electronics}

\autoref{extfig:circuit} shows the measurement circuit for the ultrasound experiment. We use a Tektronix TSG 4106A signal generator to generate RF pulses. We set the pulse position, width, and repetition frequency by external pulse modulation of the RF source supplied by a Tektronix AFG 3100 arbitrary waveform generator. Before arriving at the drive transducer the RF pulse is amplified by a Mini Circuits ZHL-42W+ power amplifier. The transmitted ultrasound excites the receive transducer, which is connected to a low noise Mini Circuits ZX60-83LN-S+ amplifier in series with a ZHL-42W+ power amplifier. We record the amplified signal from the receive transducer on a Tektronix MSO 6 Series oscilloscope. We use Mini Circuits ZFWA2-63DR+ switches to isolate the RF source and pulse amplifier from the oscilloscope. The switch logic is controlled by the AFG 3100 waveform generator. 

\subsection{Extracting amplitude and speed of sound}
Our raw data at each value of temperature and applied magnetic field is the voltage across the receive transducer as a function of time. Each voltage trace consists of a series of RF pulses---echoes---separated by the transit time of a strain pulse through the sample, typically $\sim$100 ns. We perform digital lock-in at the drive frequency to extract the relative change in the speed of sound and transmitted ultrasound amplitude. 

We measure the change in phase difference between two echoes as a function of temperature or applied magnetic field to track the change in speed of sound. At drive frequency $f$, the total phase accumulated as the sound wave travels for a time $t$ is $\phi = 2\pi f t$. The fractional change in the phase, $\delta\phi/\phi$ is then related to the fractional change in the phase velocity, $\delta v/v$, of the ultrasound by
\begin{equation}\label{sound-speed}
    \frac{\delta \phi}{\phi} = \frac{\delta t}{t} = -\frac{\delta v}{v}.
\end{equation}

The Faraday rotation is related to the amplitude of the received sound. Because each transducer is a polarization-specific detector, a change in the polarization direction of incoming transverse sound results in a change in amplitude of the voltage at the receive transducer. The amplitude of the received signal goes as  $\left|\cos(\theta)\right|$, where $\theta$ is the angle between the transducer polarization and the incoming sound wave polarization.

Changes in signal amplitude as a function of temperature and magnetic field can also result from effects unrelated to polarization rotation. Energy is lost due to ultrasound attenuation, which can change dramatically e.g. across a magnetic phase transition. Additionally, experimental artifacts can cause changes in the received amplitude: loss of collimation of the strain pulse during propagation results in interference which changes with the wavelength of the ultrasound. These interference effects can also be pronounced when there are large changes in the speed of sound, for example across a phase transition. Crucially for our experiment, these effects are all time-reversal-\textit{even}---they remain the same whether the magnetic field is up or down.

To detect the presence of time-reversal-\textit{odd} polarization rotation (i.e. Faraday rotation), we antisymmetrize the received amplitude with respect to the applied magnetic field. In addition to isolating the time-odd contribution to the signal, this process gives us several checks to confirm the consistency of the experimental protocol, described below.

\section{Data analysis---isolating the Faraday rotation}

\subsection{Antisymmetrization: transmission/reflection and compression/shear}

Our experimental protocol---45$^{\circ}$ misaligned transducer polarizations, antisymmetrizing with respect to magnetic field direction, and swapping ultrasound propagation direction---aims to isolate the antisymmetric viscous contribution to the signal. The experimental protocol also has two built-in checks to verify that the antisymmetric signal is not artificial. First, transmitted \textit{longitudinal} sound should have no antisymmetric component. Second, the antisymmetric component of the signal should disappear for both longitudinal and transverse ultrasound if the experiment is run in reflection mode where a single transducer is used for both exciting and detecting sound waves. This occurs because, even though the Faraday rotation angle accumulates over multiple reflections (and does not ``unwind" when the propagation direction is reversed), a single transducer used to excite and detect ultrasound cannot distinguish between clockwise and counterclockwise polarization rotations.

\autoref{extfig:null-longitudinal-reflection} shows the amplitude of transmitted longitudinal sound and reflected transverse sound, antisymmetrized in magnetic field for two magnetic field orientations. In both cases, within our resolution, only transmitted, transverse sound contains a field-antisymmetric component. This result is consistent with Faraday rotation and validates our method for isolating the antisymmetric signal due to Hall viscosity.

\subsection{Angle dependence}

As explained below, we expect the Hall viscosity component we are sensitive to---$\eta_{xzyz}$---to vanish with the magnetic field applied entirely in the honeycomb plane. \autoref{extfig:null-longitudinal-reflection}cd shows a comparison of the antisymmetric-in-field signal for magnetic field in the honeycomb plane compared with magnetic field with a component along the $c$ axis. Within our experimental precision, we find no antisymmetric signal due to viscosity with magnetic field entirely in the honeycomb plane.

Our claim that the Hall viscosity of \rcl is related to the coupling of phonons to spins is largely based on the observation that the Hall viscosity in our samples peaks near the critical magnetic field $B_c$. \autoref{extfig:field-angle-dependence}a shows that the peak in the Hall viscosity follows the critical magnetic field at two different tilt angles.

\subsection{Frequency dependence}

\autoref{extfig:field-angle-dependence}b shows the antisymmetric amplitude in sample 1 (210$~\mu$m thickness) measured at several frequencies with the applied magnetic field tiled 55$^\circ$ away from the $c$-axis. We find no appreciable frequency dependence over our limited transducer bandwidth for this sample. Additionally, the analysis of these data in a tilted magnetic are complicated by acoustic birefringence, which results in elliptically polarized normal modes.  

To investigate the frequency dependence further, we measured the the acoustic Faraday rotation in a fourth sample, S4, with the magnetic field purely along the $c$-axis, using a transducer with significantly larger bandwidth.  

\autoref{extfig:asym-frequency} plots the antisymmetric in magnetic field signal versus frequency squared at a fixed magnetic field strength of 12 T at 4.5 K. To obtain these data, the temperature and applied magnetic field were held fixed and the echo pattern was averaged for approximately five minutes at each frequency. The amplitude of the transmitted transverse sound wave was then antisymmetrized in field. The measurement was then repeated for the opposite propagation direction. We find that the magnitude of the antisymmetric signal scales super-linearly with frequency, though the relatively narrow frequency range prevents us from extracting a precise power law. 

However, assuming a frequency-independent Hall viscosity, the antisymmetric signal is predicted to increase as the frequency squared. In particular, solving the wave equation in \rcl with magnetic field applied along the $c$ axis and $\eta_{\text{H}}\neq0$ gives two circular polarized modes with wave vectors
\begin{equation}
	k_{\pm} = \sqrt{\frac{\rho}{c_{44} \pm \eta_{\text{H}}\omega }}\omega \approx \left(1 \mp \frac{\eta_{\text{H}}\omega}{2c_{44}}\right)\frac{\omega}{v_{44}},
\end{equation}
where $\omega$ is the normal mode frequency, $\rho$ the material density, $c_{44}$ is the elastic modulus, and $v_{44} = \sqrt{c_{44}/\rho}$ is the speed of the transverse wave. The acoustic Faraday rotation angle, $\theta$ accumulated over a distance $L$ is then
\begin{equation}
	\frac{\theta}{L} = \frac{k_+ - k_-}{2} = \frac{\eta_{\text{H}}\omega^2}{2v_{44}c_{44}}. 
\end{equation}
Therefore, when plotted against frequency squared, $f^2 = \omega^2/4\pi^2$, a linear fit to the normalized antisymmetric signal has slope 

\begin{equation}
	\frac{2\pi^2L}{v_{44}c_{44}}\eta_{\text{H}},
\end{equation}
which allows us to estimate the magnitude of \etah by a second method, independent of the one presented in the main text. 

The frequency dependence in \autoref{extfig:asym-frequency} yields \etah$\approx1.5\cdot 10^{-5}$ Pa$\cdot$s, consistent with the wave-equation fits shown in the main text. This further demonstrates the consistency of the result, both between samples (note that these frequency dependent data were taken on a different sample form those presented in the main text) and between methods used to extract the Hall viscosity. 

\subsection{Sign of the Hall viscosity}

Careful consideration of the wave equation and of our experimental geometry allows us to determine the absolute sign of the Hall viscosity tensor element that we measure---\etah. 

For magnetic field applied along the crystal $c$ axis, we have $c_{44} = c_{55}$, and the wave equation for transverse sound waves at frequency $\omega$ (Equation 2 of the main text) admits wave vectors
\begin{equation}
	\label{eqn:eignval}
	\frac{\rho\omega^2}{k_{\pm}^2} = c_{44}\pm\eta_{\text{H}}\omega,
\end{equation}
with corresponding polarizations
\begin{equation}
	\label{eqn:eignvec}
	A_{\pm} = \frac{1}{\sqrt{2}}\begin{bmatrix}\pm i \\ 1\end{bmatrix}.
\end{equation}
A sound wave launched with linear polarization, $A = A_+ + A_{-}$, initially polarized along the $y$ direction, it will decompose into normal modes that propagate as 
\begin{equation}
	\label{eqn:lin}
	A(z, t) = \frac{1}{\sqrt{2}}e^{i\omega t}\begin{bmatrix}i\left(e^{-ik_+z} - e^{-ik_-z}\right) \\ e^{-ik_+z} + e^{-ik_-z}\end{bmatrix}.
\end{equation}

Defining
\begin{equation}
	q_0 = \frac{k_+ + k_-}{2} \hspace{0.5in} \delta = \frac{k_+ - k_-}{2},
\end{equation}
we can express \autoref{eqn:lin} as 
\begin{equation}
	A(z, t) = \frac{1}{\sqrt{2}}e^{i(\omega t - q_0z)}\begin{bmatrix}i\left(e^{-i\delta z} - e^{i\delta  z}\right) \\ e^{-i\delta z} + e^{-i\delta z}\end{bmatrix} = \frac{2}{\sqrt{2}}e^{i(\omega t - q_0z)}\begin{bmatrix}\sin(\delta z) \\ \cos(\delta z)\end{bmatrix}.
\end{equation}
From \autoref{eqn:eignval}, we know that
\begin{equation}
	2\delta = \sqrt{\frac{\rho}{c_{44}}}\omega\left(\frac{1}{\sqrt{1 + \eta_{\text{H}} \omega /c_{44}}} - \frac{1}{\sqrt{1 - \eta_{\text{H}} \omega /c_{44}}}\right),
\end{equation}
so $\delta$ has the opposite sign as $\eta$ (this is generally true, and for $\eta_{\text{H}} \ll c_{44}\omega$, as in our experiment, $2\delta \approx -\eta_{\text{H}}\omega^2/v_{44}c_{44}$, where $v_{44} = \sqrt{c_{44}/\rho}$ is the speed of sound). So if $\eta_{\text{H}} >0$, as the wave moves in the positive $z$ direction the polarization rotates counter clockwise when looking at the $x$-$y$ plane from above. Equivalently, the polarization rotates clockwise when $\eta>0$ and we look down the positive $z$-axis.

To determine the sign of \etah, we keep track of the directions of the transducer polarizations on each side of the sample (i.e. in what sense they are 45$^\circ$ misaligned), the direction of the magnetic field, and which coaxial cable is connected to which transducer. Then, the sign of the antisymmetric in magnetic field amplitude can be related to the handedness of polarization rotation -- and therefore the sign of \etah-- by considering whether the sound polarization rotates into or away from the receiving transducer for a given direction of applied magnetic field.

We have carefully recorded each aspect of the experimental configuration and measured the direction of the applied magnetic field with a gaussmeter. Comparison with the analysis given above shows that with the magnetic field applied along the $c$ axis, the Hall viscosity is positive: $\eta_{xzyz}>0$.

\subsection{Comparison of symmetric and antisymmetric signals}

\autoref{extfig:data-sym-asym} shows both the symmetric and antisymmetric in magnetic field signals for the magnetic field applied along $c$ and tiled 55$^\circ$ away from $c$. Note that the antisymmetric signal has a distinct magnetic field dependence from the symmetric signal, indicating that our antisymmetrization procedure separates components of the signal due to distinct mechanisms. While the symmetric, reciprocal, data are due to even-in-magnetic-field processes such as ultrasound attenuation, the antisymmetric, non-reciprocal data are due to Hall viscosity. 

\subsection{Value of the Faraday rotation angle}

\autoref{extfig:faraday-angle-field} plots the antisymmetric signal converted to a polarization rotation angle. In particular, consider a linear polarized transverse whose polarization has rotated by an angle $\theta$ about the propagation direction when the magnetic field is greater than zero. When incident on a transducer with polarization misaligned 45$^\circ$ from the drive transducer polarization, the amplitude will be $A(B>0) = A_0 |\cos(\pi/4+\theta)|$, where $A_0$ is the even-in magnetic field amplitude. When the direction of the applied magnetic field is reversed, the amplitude will be $A(B<0) = A_0 |\cos(\pi/4-\theta)|$. Therefore, when the rotation angle is less than $\pi/4$ (as in our data), the rotation angle can be related to the ratio of the symmetric and antisymmetric data as

\begin{equation}
\frac{A(B>0) - A(B<0)}{A(B>0) + A(B<0)} = \frac{A_{\text{asym}}}{A_{\text{sym}}} = \tan^{-1}(\theta).
\end{equation}

\subsection{Antisymmetric Signal at Elevated Temperatures}

We find that the Hall viscosity persists to temperatures above the N\'{e}el temperature of $T_{\rm{N}} \approx 7.5$ K. \autoref{extfig:high-temp-asym} shows the antisymmetric-in-field data used to estimate the Hall viscosity at temperatures well inside the antiferromagnetic phase, just below the N\'{e}el transition, and well above the N\'{e}el transition in the paramagnetic phase.\\

\section{Data analysis -- estimating the viscosity}

\subsection{Wave equation with viscosity and birefringence}
To estimate the magnitude of the Hall viscosity from our pulse echo data, we use the wave equation for transverse sound waves propagating along the $c$ axis of \rcl in the presence of Hall viscosity. The Lagrangian for small displacements $\vec{u}$ is
\begin{equation}\label{eq:gen-L}
    \mathcal{L} = \frac{\rho}{2}\dot{u}^2 - \frac{c_{ijkl}}{2}\varepsilon_{ij}\varepsilon_{kl} - \frac{\eta_{ijkl}}{2}\varepsilon_{ij}\dot{\varepsilon}_{kl},
\end{equation}
where $\rho$ is the material density, $\varepsilon_{ij} = (\partial_iu_j + \partial_ju_i)/2$ are the lattice strains, $c_{ijkl}$ are the elastic moduli, and $\eta_{ijkl}$ are the antisymmetric Hall viscosities. For harmonic displacements with wave vector $k$ along the $c$ axis, and with Hall viscosity $\eta\equiv \eta_{xzyz}$ to which we are sensitive in our experiment, \autoref{eq:gen-L} becomes
\begin{equation}\label{eq:exp-L}
\mathcal{L} = \frac{\rho}{2}\dot{u}^2 - \frac{c_{44}}{2}k^2u_y^2 - \frac{c_{55}}{2}k^2u_x^2 - \frac{\eta}{2}k^2(u_x\dot{u}_y - \dot{u}_xu_y),
\end{equation}
from which we derive the equation of motion presented in the main text:

\begin{equation}\label{EOM}
\rho\omega^2 \vec{u} = k^2\begin{bmatrix} c_{55} & i\eta \omega \\ -i\omega \eta & c_{44} \end{bmatrix}\vec{u}.
\end{equation}
Note that, while $c_{44} = c_{55}$ in zero magnetic field for \rcl, this will no longer be the case with a magnetic field component in the honeycomb plane. This in-plane magnetic field dependence between $c_{44}$ and $c_{55}$ is known as acoustic birefringence and, crucially, is even in magnetic field (time reversal even). 

We can gain some intuition for the equation of motion by rewriting the matrix on the right-hand side in terms of the average elastic modulus $c = (c_{44}+c_{55})/2$ and the difference $\delta = (c_{44}-c_{55})/2$. Then the equation of motion becomes a sum of Pauli matrices,
\begin{equation}\label{eq:EOM-pauli}
\begin{bmatrix} c_{55} & i\eta \omega \\ -i\omega \eta & c_{44}. \end{bmatrix} = c1 + \delta \sigma_z + \omega\eta\sigma_y
\end{equation}
The square of the phase velocities, $v_{\pm}^2 = \omega^2/k^2$, are then
\begin{equation}\label{eq:v-ph}
\rho v_{\pm}^2 = c \pm \sqrt{\delta^2 + \omega^2\eta^2},
\end{equation}
so that the two transverse wave speeds are split both by the difference in elastic moduli and by the viscosity. The polarizations are, up to a normalization constant,
\begin{equation}\label{eq:pol}
\mathbf{u}_+ = \frac{1}{\sqrt{\delta^2 + \omega^2\eta^2}}\begin{bmatrix} \delta + \sqrt{\delta^2 + \omega^2\eta^2} \\ i\omega \eta \end{bmatrix} \hspace{0.25in} \mathbf{u}_- = \frac{1}{\sqrt{\delta^2 + \omega^2\eta^2}}\begin{bmatrix} i\omega \eta \\ \delta + \sqrt{\delta^2 + \omega^2\eta^2}\end{bmatrix}. 
\end{equation}
From \autoref{eq:pol} we can see that when $\eta = 0$, changing $\delta$ does not change the polarizations thought it does split the degenerate speeds of sound. When $\delta = 0$ adding non-zero viscosity results in circular rather than linear polarizations. Finally, when both $\eta, \delta \neq 0$, we get elliptical polarizations.

\subsection{Wave equation fits to the data}

We use the equations of motion---\autoref{EOM}---to model the ultrasound experiment. Our experiment consists of launching a strain wave along the $c$ axis with frequency $\omega$ and polarization at angle $\phi_i$ in the $x$-$y$ plane: $\mathbf{u}_i = (\cos(\phi_i), \sin(\phi_i))$. To obtain the amplitude that we measure at the receive transducer for a given value of the viscosity, we proceed in three steps.

First, we compute the normal mode wave vectors, $k _{\pm}$, and polarizations $\mathbf{u}_\pm$, using \autoref{EOM}. The initial polarization is then decomposed into the normal modes:
\begin{equation}\label{polar-decomp}
\mathbf{u}_i = \alpha \mathbf{u}_+ + \beta\mathbf{u}_-,
\end{equation}

from which we can find the coefficients $\alpha = \mathbf{u}_+^{\dagger}\cdot\mathbf{u}_i$ and $\beta = \mathbf{u}_i^{\dagger}\cdot\mathbf{u}_i$. The phase of each mode after traveling the length of the sample, $\ell$, is $e^{ik_{\pm}\ell}$. We then project each mode on to the polarization  of the receive transducer $\mathbf{u}_f = (\cos(\phi_f), \sin(\phi_f))$, giving two contributions to the final (complex) amplitude, $\tilde{A}_{\pm}$  from each mode polarization $\mathbf{u}_{\pm}$

\begin{equation}\label{polar-comp}
\tilde{A}_{\pm} = \left(\mathbf{u}_{\pm}^{\dagger}\cdot\mathbf{u}_i\right)\left(\mathbf{u}_{f}^{\dagger}\cdot\mathbf{u}_{\pm}\right)e^{ik_{\pm}\ell}
\end{equation}

The receive transducer sees the sum of the two modes, and the amplitude we measure, $A_m$ is the amplitude of the sum

\begin{equation}\label{amp-meas}
A_m = \left|\tilde{A}_+ + \tilde{A}_-\right|
\end{equation}

By flipping the sign of $\eta$ in the model we form symmetrized and antisymmetrized amplitudes as a function of applied magnetic field. Because the antisymmetric amplitude alone is not a meaningful quantity in this context (it depends on the overall amplitude of the initial excitation, for example) we normalize the antisymmetric amplitude in the model and the data by the symmetric amplitude at zero applied magnetic field. To extract the viscosity as a function of applied magnetic field, we perform a one parameter fit to the ratio of the antisymmetric and zero field symmetric amplitudes using the model presented above. The value of viscosity as a function of magnetic field presented in the main text minimizes the squared difference between the model output and the data.\\

\subsection{Hall viscosity tensor in \rcl} 
\label{subsec:group-theory}
The viscosity tensor, $\eta$ relates stress to the time rate of strain $\dot{\varepsilon}$ and gives a contribution to the elastic energy density
\begin{equation}\label{stress-strain}
    U_{\text{el}} = \frac{c_{ijkl}}{2}\varepsilon_{ij}\varepsilon_{kl}+\frac{\eta_{ijkl}}{2}\varepsilon_{ij}\dot{\varepsilon}_{kl},\end{equation}
where $c$ is the elastic tensor. 

Because strain is a symmetric tensor, the viscosity is also symmetric under interchange within the first and second pairs of indices: $i \longleftrightarrow j$ and $k \longleftrightarrow l$. It is also clear from \autoref{stress-strain} that the viscosity is odd under time reversal.

The viscous contribution to the energy comes from the product of two strains. The viscosity tensor, then, can be divided into symmetric and antisymmetric parts under exchange of those strains -- the  first and second pair of indices $ij \longleftrightarrow kl$. The symmetric viscosity, $\eta^S = (\eta_{ijkl} + \eta_{klij})/2$, results in energy loss and is even under reversing the sign of an applied magnetic field. In the context of an ultrasound experiment $\eta^S$ is proportional to the ultrasonic attenuation. The antisymmetric, Hall viscosity, $\eta^H = (\eta_{ijkl} - \eta_{klij})/2$, generates acoustic Faraday rotation and an intrinsic thermal Hall effect.

When setting out to measure the Hall viscosity, it is important to know when it is required to vanish by symmetry. Counting the symmetry-allowed components of $\eta^H$ is simplified by working with the irreducible representations of strain. The elastic energy due to the Hall viscosity coupling two irreducible strains $\varepsilon_\Gamma$ and $\varepsilon_{\Gamma'}$ is
\begin{equation} \label{eq:Uel}
U_{\text{visc}} = \frac{\eta_{\Gamma\Gamma'}}{2}(\dot{\varepsilon_{\Gamma}}\varepsilon_{\Gamma'} - \varepsilon_{\Gamma}\dot{\varepsilon}_{\Gamma'}).
\end{equation}

The system, assumed to break time reversal symmetry (whether intrinsically or through an applied magnetic field), allows $\eta_{\Gamma\Gamma'}$ to be time odd. Both the energy and tensor element $\eta_{\Gamma\Gamma'}$ are scalars ($A_{g}$ objects in the $\rm{S}_6$ point group), which requires the term in parentheses in \autoref{eq:Uel} to also be a scalar. Because $\eta_{\Gamma\Gamma'}$ is antisymmetric under exchange $\Gamma \longleftrightarrow \Gamma'$, this forces either: 1) that $\varepsilon_{\Gamma} = \varepsilon_{\Gamma'}$ and that the representation $\Gamma$ squared contains an antisymmetric scalar (and therefore must be two or three dimensional); or 2) that $\varepsilon_{\Gamma}$ and $\varepsilon_{\Gamma'}$ are distinct strains that belong to the same representation.

When time reversal symmetry is broken by an applied magnetic field, it is more transparent to ``factor out'' the magnetic field from the Hall viscosity. Then, to identify which magnetic field components generate which Hall viscosity coefficients, we expand the viscosity to linear order in applied magnetic field. The magnetic field splits into representations, $B_\Sigma$, and we form the antisymmetric products which each contribute to the elastic energy,
\begin{equation}\label{lin-visc}
U_{\text{visc}}(B_\Sigma) = B_\Sigma~\tilde{\eta}_{\Gamma\Gamma'}^\Sigma(\dot{\varepsilon_{\Gamma}}\varepsilon_{\Gamma'} - \varepsilon_{\Gamma}\dot{\varepsilon}_{\Gamma'}),
\end{equation}
where now the coefficient $\tilde{\eta}_{\Gamma\Gamma'}^\Sigma$ is a scalar and has unit of Pa$\cdot$s$\cdot$T$^{-1}$, and the antisymmetric term in parentheses transforms as $\Sigma$ so that the whole expression is a scalar. Note that the product $\dot{\varepsilon_{\Gamma}}\varepsilon_{\Gamma'}$ generically decomposes into the sum of multiple representations, not all of which transform like $\Sigma$. To avoid cluttering the notation further, we specify which portion of the product is involved in a particular term with the superscript on the coefficient $\tilde{\eta}^{\Sigma}_{\Gamma\Gamma'}$ (but again note that $\tilde{\eta}^{\Sigma}_{\Gamma\Gamma'}$ itself is always a scalar).

Now we take the specific example of \rcl. Below its structural transition at 150 K, \rcl transitions to rhombohedral symmetry with point group S$_6$. There are four irreducible strains in S$_6$: two $A_g$ strains and two $E_g$ strains:
\begin{equation*}\label{s6-irreps}
\begin{aligned}
&\varepsilon_{A_g^{(1)}} = \varepsilon_{x^2 + y^2}, ~&\varepsilon_{A_g^{(2)}} = \varepsilon_{z^2},
~~&\varepsilon_{E_g^{(1)}} = \left\{\varepsilon_{x^2 - y^2},\varepsilon_{xy}\right\}, ~&\varepsilon_{E_g^{(2)}} = \left\{\varepsilon_{xz},\varepsilon_{yz}\right\}
\end{aligned}.
\end{equation*}

The magnetic field splits into two representations: the out-of-plane component transforms as $A_g$, and the in-plane component transforms as $E_g$:
\begin{equation*}\label{bfield-s6-irreps}
\begin{aligned}
&B_{A_g} = B_z, ~&B_{E_g} = \left\{B_x, B_y\right\}
\end{aligned}.
\end{equation*}

To simplify the notation in what follows, we drop the $g$ in the representation labels since we only need to consider representations that are even under inversion. As mentioned earlier, because the product of two strains can decompose into the sum of several other strains, we use the notation from \autoref{lin-visc} to specify which viscosity coefficient goes with which strain in the sum. For example, one such term will be
\begin{equation}\label{sample}
U_{\text{visc}}(B_{A_{1g}}) \sim B_{A_{g}}\tilde{\eta}^{A_{g}}_{E_{g}^{(1)}E_{g}^{(2)}}(\asym{E_g^{(1)}}{E_g^{(2)}}) \equiv B_{A}\tilde{\eta}^{A}_{E_1 E_2}(\asym{E_1}{E_2}),
\end{equation}
where  $\tilde{\eta}^{A}_{E_1E_2}$ denotes a viscosity term that couples the $A_{g}$ portion of the product of $E_{g}^{(1)}$ and $E_{g}^{(2)}$ strains to the $B_{A_{g}}$ magnetic field component.

With an out-of-plane magnetic field $B_z$, we can form the following four Hall viscosity contributions to the elastic energy:
\begin{align}
U_{\text{visc}}(B_z) = B_{A} \left[
\tilde{\eta}^{A}_{A_1A_2}(\asym{A_1}{A_2})
+ \tilde{\eta}^{A}_{E_1E_2}(\asym{E_1}{E_2})
\right. \notag \\
\left.
+ \tilde{\eta}^{A}_{E_1E_1}(\asym{E_1}{E_1})
+ \tilde{\eta}^{A}_{E_2E_2}(\asym{E_2}{E_2})
\right]\label{bz-visc},
\end{align}
where the last two terms, which involve the square of a representation, arise because the product of $E_g$ with itself contains an antisymmetric object which transforms as $A_g$.

An in-plane magnetic field generates five Hall viscosity terms:
\begin{align}
U_{\text{visc}}(B_x, B_y) = B_{E} \left[
\tilde{\eta}^E_{A_1E_1}(\asym{A_1}{E_1})
+ \tilde{\eta}^E_{A_1E_2}(\asym{A_1}{E_2})
\right. \notag \\
\left.
\tilde{\eta}^E_{A_2E_1}(\asym{A_2}{E_1})
+ \tilde{\eta}^E_{A_2E_2}(\asym{A_2}{E_2})
\right. \notag \\
\left.
+ \tilde{\eta}^E_{E_1E_2}(\asym{E_1}{E_2})
\right]\label{bxby-visc}.
\end{align}

The Hall viscosity we measure in our experiment shows up here as $B_A\tilde{\eta}^{A}_{E_2E_2} = \eta_{xzyz}$ and is only generated by the out-of-plane magnetic field, $B_z$. When the magnetic field is tilted away from the $c$ axis (but with some component still along $z$), all nine Hall viscosity coefficients from both \autoref{bz-visc} and \autoref{bxby-visc} can be generated, including $\eta_{xzyz}$.

\subsection{Heat flux in the presence of Hall viscosity}
To estimate the intrinsic contribution to the thermal Hall effect due to Hall viscosity, we consider the transverse energy flux generated by thermal phonons. Thermally generated stress at frequency $\omega$, $\sigma(\omega)$, carries an acoustic energy current
\begin{equation}\label{en-flux}
    j_i = \text{Re}[\sigma_{ij}(\omega)\dot{u}_{j}] = \text{Re}[\left(c_{ijkl}\varepsilon_{ij} + \eta_{ijkl}\dot{\varepsilon}_{kl}\right)\dot{u}_{j}],
\end{equation}
where $\dot{u}$ is the time derivative of displacements due to the stress \citemethods{loveElasticity}. Longitudinal thermal transport occurs because temperature gradients produce gradients in the amplitude (energy occupations) of thermal phonons. For example, if we consider $z$ polarized thermal strains propagating in the $x$-direction with wave vector $q_x$, so that $u = u_z(x)e^{i(\omega t - q_x x)}$, there is a contribution to the energy flux
\begin{equation}\label{kxx-j}
j_x = \text{Re}[\sigma_{xz}\dot{u}_{z}] = \text{Re}[i\omega c_{55}\varepsilon_{xz}u_z] = c_{55}\omega  q_x u_z^2 ,
\end{equation}
where we use Hooke's law $\sigma_{xz} = iq_x c_{55}u_z$ for this particular component. Strains involving displacements along $x$ and $y$ also contribute to $j_x$, bringing in terms proportional to $c_{11}$ and $c_{66}$ respectively. Taken together we can combine these contributions, using an average elastic modulus $\overline{c}$, to obtain
\begin{equation}\label{kxx-jx}
    j_x = \overline{c}~ q_x\omega~u^2 = \frac{\overline{c}}{\rho v_s} \left(\rho \omega^2u^2\right) = v_s \left(\rho \omega^2u^2\right),
\end{equation}
where $v_s$ is an average speed of sound and $\rho$ is the density. Because these are thermally generated, random strains, there is no preferred direction for $q_x$ and, in the absence of a thermal gradient, the net energy flux is zero. \\

However, in the presence of a thermal gradient along $x$, there is heat transport. Conceptually, the thermal phonons travel a mean free distance $\ell$ before they scatter and equilibrate with the local lattice temperature. The net acoustic energy flux is then
\begin{equation}
j_{x, \text{net}} = j_x(x + \ell) - j_x(x - \ell) = v_s \left(\rho \omega^2u(x+\ell)^2 - \rho \omega^2u(x-\ell)^2\right).
\end{equation}
Recognizing that $\rho \omega^2 u^2$ is the phonon energy density, $U$, we write
\begin{equation}\label{jx-net}
j_{x, \text{net}} \approx v_s \ell \frac{\partial U}{\partial x} = v_s \ell C \frac{\partial T}{\partial x} = \kappa_{xx}\frac{\partial T}{\partial x}
\end{equation}
where $C$ is the heat capacity, and we identify the longitudinal thermal conductivity $\kappa_{xx} = v_s \ell C$. This thermal conductivity is extrinsic in the sense that it is proportional to the phonon mean free path.

Now consider the heat flux when there is a non-zero Hall viscosity. In particular, we  consider an applied magnetic field along the $c$ axis of \rcl. Returning to \autoref{en-flux}, we find that the Hall viscosity produces a transverse energy flux where, for example, the viscosity $\eta_{xzyz}$ produces
\begin{equation}
j_y = \text{Re}[2\eta_{xzyz} \dot{\varepsilon}_{xz}\dot{u}_z] = \eta_{xzyz} \omega^2\frac{\partial u_z}{\partial x}u_z = \frac{\eta_{xzyz}}{2\rho} \frac{\partial}{\partial x}\left(\rho \omega^2 u_z^2\right),
\end{equation}
and the other two viscosity components $\eta_{yyxy}$ and $\eta_{yxxx}$ contribute similar terms. In the absence of a temperature gradient, there will be no net transverse heat flow, but if we enforce an amplitude gradient by applying a temperature gradient along the $x$-direction, then we transfer the spatial derivative to the applied temperature gradient and write
\begin{equation}\label{trans-j}
    j_y = \frac{\overline{\eta}}{\rho}C\frac{\partial T}{\partial x},
\end{equation}
where $\overline{\eta}$ is an average Hall viscosity. Note that, while the transverse energy flux requires a longitudinal temperature gradient, its magnitude does not depend on the phonon mean free path---this effect is intrinsic.\\

Finally, the thermal Hall angle---the ratio of the thermal Hall conductivity and longitudinal thermal conductivity---is equal to the ratio of the transverse and longitudinal heat currents. Using \autoref{jx-net} and \autoref{trans-j} we recover the estimate
\begin{equation}\label{kxx-kxy}
\frac{\kappa_{xy}}{\kappa_{xx}} = \frac{\eta}{\rho}\frac{C}{\kappa_{xx}}.
\end{equation}
Note that using this equation to compare our measurement of $\eta_{xzyz}$ to the thermal Hall effect depends on the Hall viscosity being only weakly dependent on frequency (since thermal measurements access much higher phonon frequencies than do our ultrasound measurements). This assumption is justified by the calculations of Dhakal \textit{et al.} \cite{dhakalTheoryIntrinsicPhonon2025}, who find that the Hall viscosity is independent of frequency for small wavevector.

\subsection{Sample dependence}

\autoref{fig:sample-dependence} shows the antisymmetric signal with the magnetic field tilted 45$^\circ$ away from the $c$ axis toward the $b$ axis for three samples, all of which have $T_N \approx 7.5$ K. We observe Faraday rotation in all three samples, finding non-zero Hall viscosity in all three samples. There is variation in the size of the antisymmetric signal between samples, and the thicker samples have additional structure near the phase transition. We attribute this, at least in part, to the likelihood of magnetic domains in the thicker samples. Note that while magnetic domains can cause structure in the antisymmetric signal, they cannot be the \textit{origin} of the antisymmetric signal.

\subsection{Sample Quality Over Time}

\autoref{fig:TN-time} shows the speed of sound across the Neel transition of sample 1 (200$ \mu$m thickness) at zero field during two measurement sequences separated by one year. We did not see any change in $T_N$ on thermally cycling our samples numerous times. 

%\bibliographystylemethods{plain}
%\bibliographymethods{literature}

\section{Acknowledgments:}
A. S. and B. J. R. acknowledge helpful discussions with L. Balents, H. Guo, G. Grissonnanche C. Jian, H. Y. Kee, Y. B. Kim, P. A. Lee, A. MacDonald, K. Modic, S. Sachdev, A. Shekhter, L. Taillefer, R. Valenti, D. Vanderbilt, and S. Winter. Work at Cornell University (sample preparation, data acquisition, dat analysis) was funded through the Air Force Office of Scientific Research under grant \# FA9550-23-1-0306, ``Phonon Berry Curvature in Quantum Materials''. Additional funding was provided by the Canadian Institute for Advanced Research. Work at the University of Toronto (sample growth and characterization) was supported by the Natural Sciences and Engineering Research Council of Canada (NSERC), [funding reference numbers RGPIN-2019-06449, RTI-2019-00809, and RGPIN-2025-06514], and by the Canada Foundation for Innovation (CFI) and the Ontario Research Fund (ORF) for Project No. 36404.

\section{Author Contributions}
B. J. R. and A.S. conceived the experiment. E.H., S.K, and Y.J.K. grew and characterized the samples of \rcl. A.S. prepared the samples for ultrasound experiments, collected and analyzed the data, and produced the figures. B.J.R. and A.S. wrote the manuscript, with input from all coauthors. B.J.R. supervised the project.

\subsection{Competing interests:} The authors declare no competing interests. 

\subsection{Data and Code availability:}
The data that support the findings of this article are openly available at \url{https://github.com/CHiLL-Ramshaw/manuscripts-supporting_data/tree/b83031e9f6dbbc36103f41f000949129f8e14573/2026_RuCl3_Viscosity}

\section{Extended Data Figures and Tables}

\begin{extfigure}[H]
	\begin{center}
		\includegraphics[width=.8\textwidth]{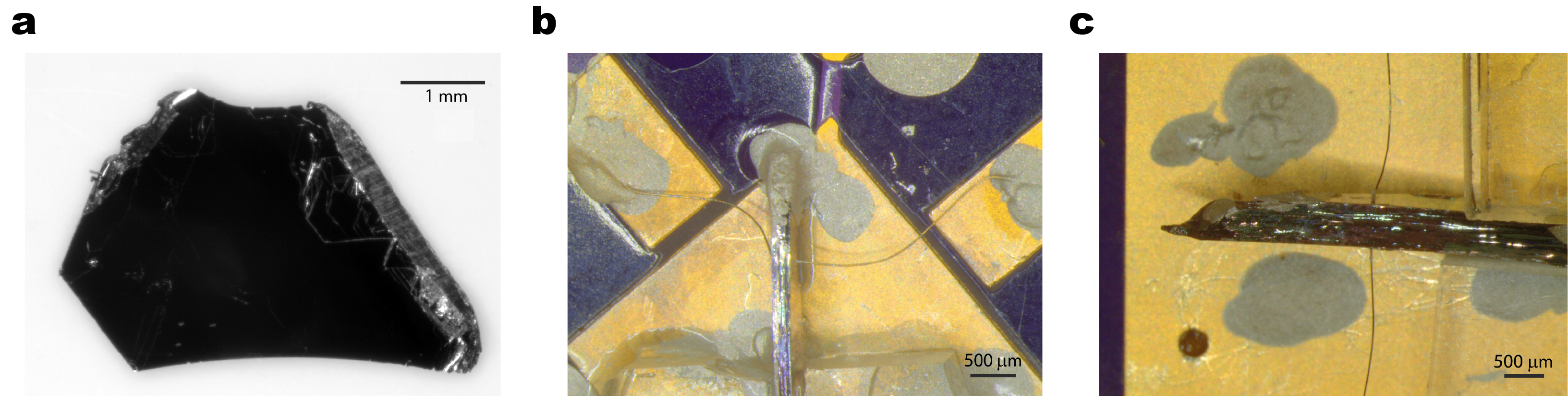}
	\end{center}
	\caption{\textbf{\rcl Samples: }\textbf{a}, shows a representative sample of \rcl used for the ultrasound experiment before the piezoelectric ZnO is grown on the sample surface. A pristine (001) face is exposed. The lateral dimensions are several millimeters and the thickness is several hundred microns. \textbf{b--c}, show samples 1 and 2 (see Extended Data Table 1) after ultrasonic transducers have been deposited on both exposed (001) surfaces of each sample, and the samples have been mounted on PCBs for the ultrasound transmission measurement. The two wires running to the sample surface are used to excite (read out) the drive (receive) transducer.}
  \label{extfig:sample-photo}
\end{extfigure}

\begin{table}[h]
  \centering
  \begin{tabular}{|p{3cm}|p{3cm}|p{3cm}|}
    \hline
    Sample Name & Thickness ($\mu$m) & $T_N$ (K) \\
    \hline
    S1 & 210 & 7.6 \\
	\hline
    S2 & 550 & 7.7 \\
	\hline
    S3 & 490 & 7.7 \\
    \hline
    S4 & 320 & 7.8 \\
  \hline
  \end{tabular}
  \caption{\textbf{Table of samples.} Sample name, the sample thickness in the $c$ direction, and the Neel temperature.}
  \label{tab:samples}
\end{table}

%\begin{spacing}{1}
%Extended Data Tab. 1\xspace\rule{0.4pt}{1em}\xspace\textbf{Table of samples:} Sample name, the sample thickness in the $c$ direction, and the Neel temperature.
%\end{spacing}

\begin{extfigure}[H]
	\begin{center}
		\includegraphics[width=.8\textwidth]{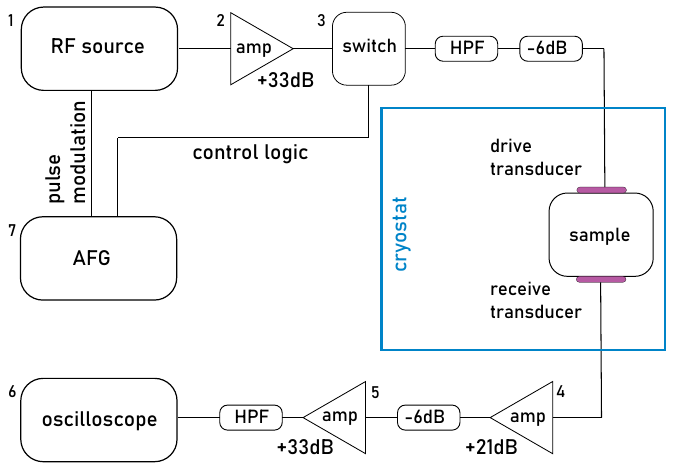}
	\end{center}
	\caption{\textbf{Measurement circuit: }Schematic of the measurement circuit, starting from the top left and moving clockwise. We generate RF pulses at the transducer drive frequency ($\sim$1 GHz) using a Tektronix TSG 4106A signal generator (1). The source is pulse modulated using a Tektronix AFG 3100 waveform generator (7) to output $\sim$30 ns square pulses at a repetition frequency of $\sim$100 kHz. The RF pulse is further amplified by a Mini Circuits ZHL-42W+ power amplifier with 33 dB gain (2).  The pulse is then fed through a Mini Circuits ZFWA2-63DR+ switch (3) to isolate the sample and downstream circuit from the power amplifier after the initial drive pulse is sent to the transducer. The switch logic is controlled by a second channel on the AFG 3100 generator (7). Before reaching the sample, the pulse is high-pass-filtered (HPFed) to remove switching noise and attenuated to minimize unwanted reflections from the transducer. The signal at the receive transducer -- the transmitted ultrasound -- is amplified twice, first by a low noise Mini Circuits ZX60-83LN-S+ (4) then by a second ZHL-42W+ power amplifier (5) before being filtered and recorded on a Tektronix MSO 6 Series oscilloscope (6).}
  \label{extfig:circuit}
\end{extfigure}

\begin{extfigure}[H]
	\begin{center}
		\includegraphics[width=.8\textwidth]{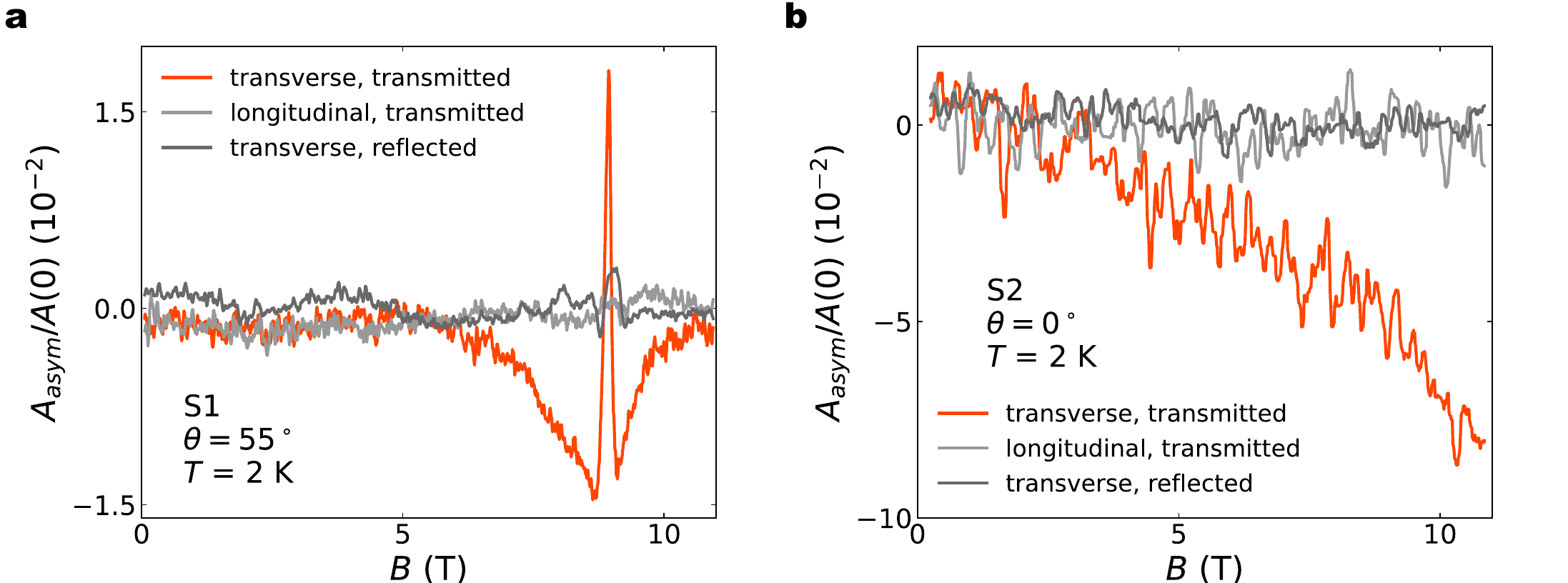}
		\includegraphics[width=.8\textwidth]{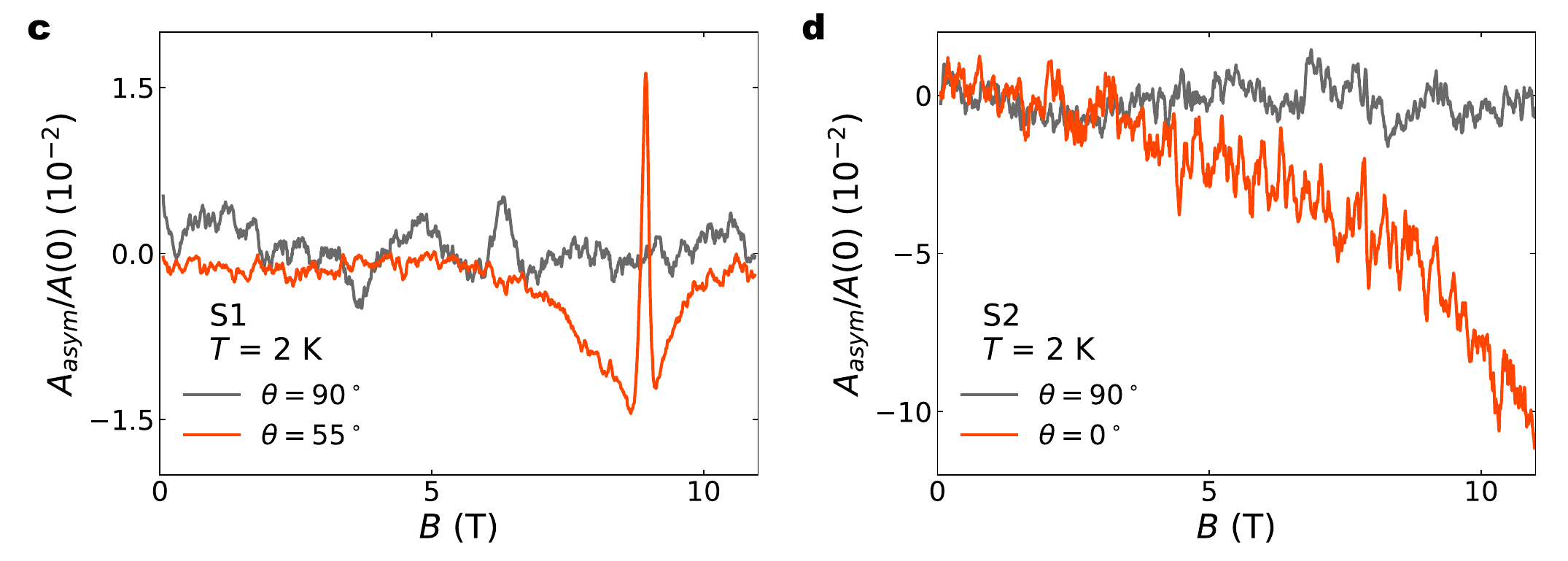}
	\end{center}
	\caption{\textbf{Null signal of longitudinal and reflected transverse sound, and null signal with an in-plane magnetic field: } To confirm that the antisymmetric signal we measure is not an artifact of the experimental technique, we show the antisymmetric in field amplitude of transmitted longitudinal sound and reflected transverse sound. Compressional sound does not exhibit a Faraday rotation and therefore should have no antisymmetric in field component. Reflected transverse sound can exhibit a Faraday rotation, but the effect disappears on antisymmetrization because when the same transducer is used to excited and detect the sound wave the transducer cannot distinguish between clockwise and counterclockwise rotations. \textbf{a}, shows the null signal for sample 1 in a tilted magnetic field and \textbf{b}, shows the null signal for sample 2 with the magnetic field along the crystal $c$ axis. In both cases, the antisymmetric signal for transmitted transverse sound exceeds the null signal by an order of magnitude. Symmetry constrains the Faraday rotation to occur only when the applied magnetic field has a component along the crystal $c$ axis. Here we show the antisymmetric signal for magnetic field applied entirely in the plane ($\theta$ = 90$^\circ$, where we expect no Faraday rotation) versus with an out-of-plane component of the magnetic field. \textbf{c}, shows the result for sample 1 along with the 55$^\circ$ data presented in the main text. \textbf{d}, shows the same for sample 2 with the 0$^\circ$ data from the main text. In both cases, we find no evidence of Faraday rotation with the magnetic field applied entirely in the honeycomb plane. Note that this does not imply that there is no thermal Hall effect for this magnetic field orientation; only that the Hall viscosity component we measure, $\eta_{xzyz}$, is zero.}
  \label{extfig:null-longitudinal-reflection}
\end{extfigure}

%\begin{extfigure}[H]
%	\begin{center}
%		\includegraphics[width=.8\textwidth]{Extended-Data-Fig-4.pdf}
%	\end{center}
%	\caption{\textbf{Null signal with in-plane magnetic field: }Symmetry constrains the Faraday rotation to occur only when the applied magnetic field has a component along the crystal $c$ axis. Here we show the antisymmetric signal for magnetic field applied entirely in the plane ($\theta$ = 90$^\circ$, where we expect no Faraday rotation) versus with an out-of-plane component of the magnetic field. \textbf{a}, shows the result for sample 1 along with the 55$^\circ$ data presented in the main text. \textbf{b}, shows the same for sample 2 with the 0$^\circ$ data from the main text. In both cases, we find no evidence of Faraday rotation with the magnetic field applied entirely in the honeycomb plane. Note that this does not imply that there is no thermal Hall effect for this magnetic field orientation; only that the Hall viscosity component we measure, $\eta_{xzyz}$, is zero.}
%  \label{extfig:null-in-plane}
%\end{extfigure}

\begin{extfigure}[H]
	\begin{center}
		\includegraphics[width=.45\textwidth]{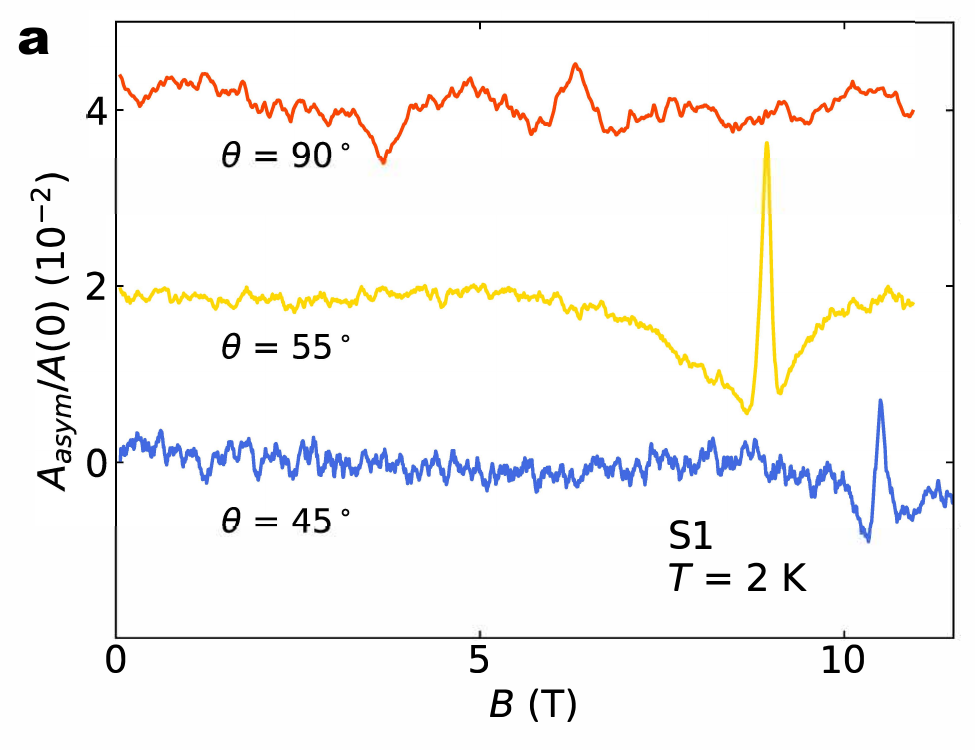}
		\includegraphics[width=.45\textwidth]{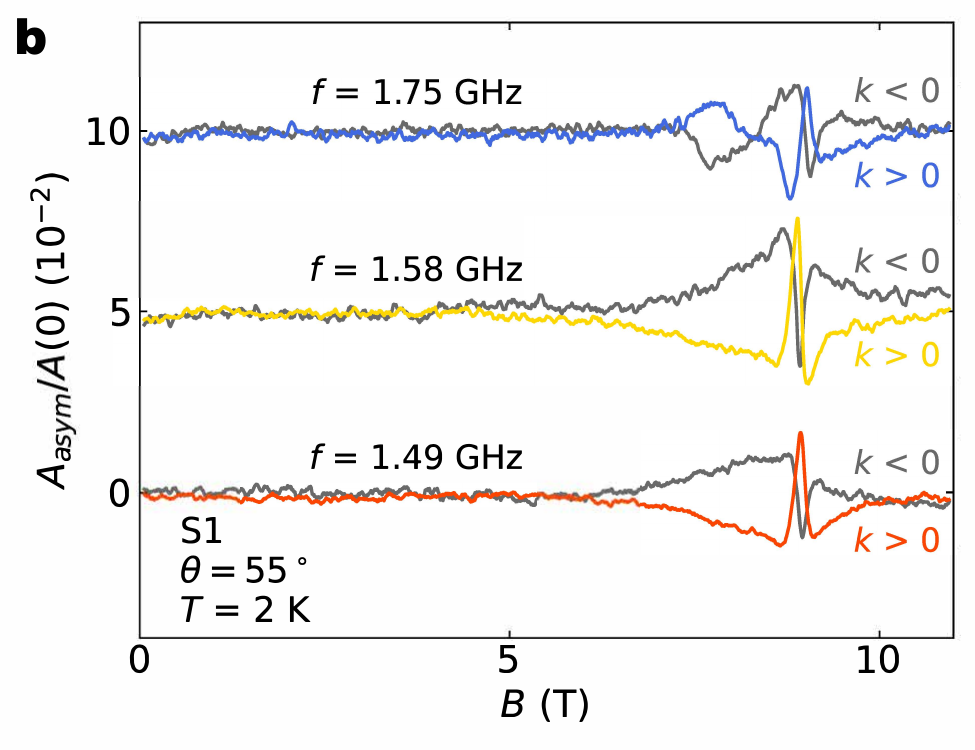}
	\end{center}
	\caption{\textbf{Magnetic field angle dependence, and frequency dependence at a fixed angle: }To investigate the origin of the Hall viscosity of \rcl we measure the Faraday rotation as a function of the out-of-plane tilt angle of the magnetic field (\textbf{a}). Here we show the antisymmetric in field amplitude for three orientations of the applied magnetic field. In both tilted field orientations the onset and peak of the antisymmetric signal -- and therefore the Hall viscosity -- tracks the critical magnetic field where zigzag AFM order is destroyed, moving to higher magnetic field as the field is tilted towards the $c$ axis. This suggests that the Hall viscosity of \rcl is connected to the spin degrees of freedom in the material. (\textbf{b}) Here we show the antisymmetric in field amplitude over a range of frequencies for both propagation directions.}
  \label{extfig:field-angle-dependence}
\end{extfigure}

%\begin{extfigure}[H]
%	\begin{center}
%		\includegraphics[width=.8\textwidth]{Extended-Data-Fig-6.pdf}
%	\end{center}
%	\caption{\textbf{Frequency dependence: }Here we show the antisymmetric in field amplitude over a range of frequencies for both propagation directions.}
%  \label{extfig:field-sweep-frequency}
%\end{extfigure}

\begin{extfigure}[H]
	\begin{center}
		\includegraphics[width=.8\textwidth]{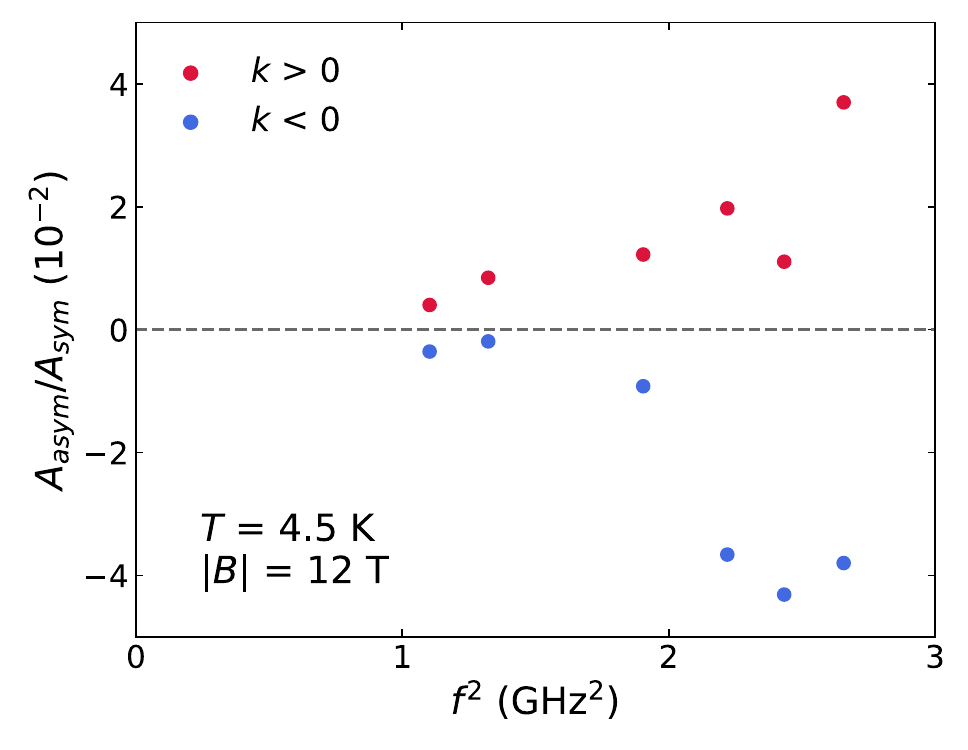}
	\end{center}
	\caption{\textbf{Frequency dependence at fixed magnetic field strength: }The antisymmetric signal measured at fixed magnetic field strength of 12 T as a function of frequency for positive and negative propagation directions. In addition to the magnetic field sweeps shown in the main text, we perform extended averaging at both positive and negative magnetic fields over the full bandwidth of the transducer. Measuring a fourth sample, S4, over a frequency span of approximately 600 MHz, we find the antisymmetric signal has a super-linear dependence on frequency. Under the assumption of a frequency independent viscosity, the Faraday rotation angle scales with the square of the ultrasound frequency, providing an independent method of estimating the Hall viscosity of \rcl. A linear fit to the frequency dependence shown here gives a Hall viscosity $\etah \approx 1.5\cdot 10^{-5}$ Pa$\cdot$s, consistent with the wave equation fits to the magnetic field sweeps plotted in the main text. Note that the data appears to grow faster than linear, which indicates that the viscosity itself grows with increasing frequency. }
  \label{extfig:asym-frequency}
\end{extfigure}

\begin{extfigure}[H]
	\begin{center}
		\includegraphics[width=.8\textwidth]{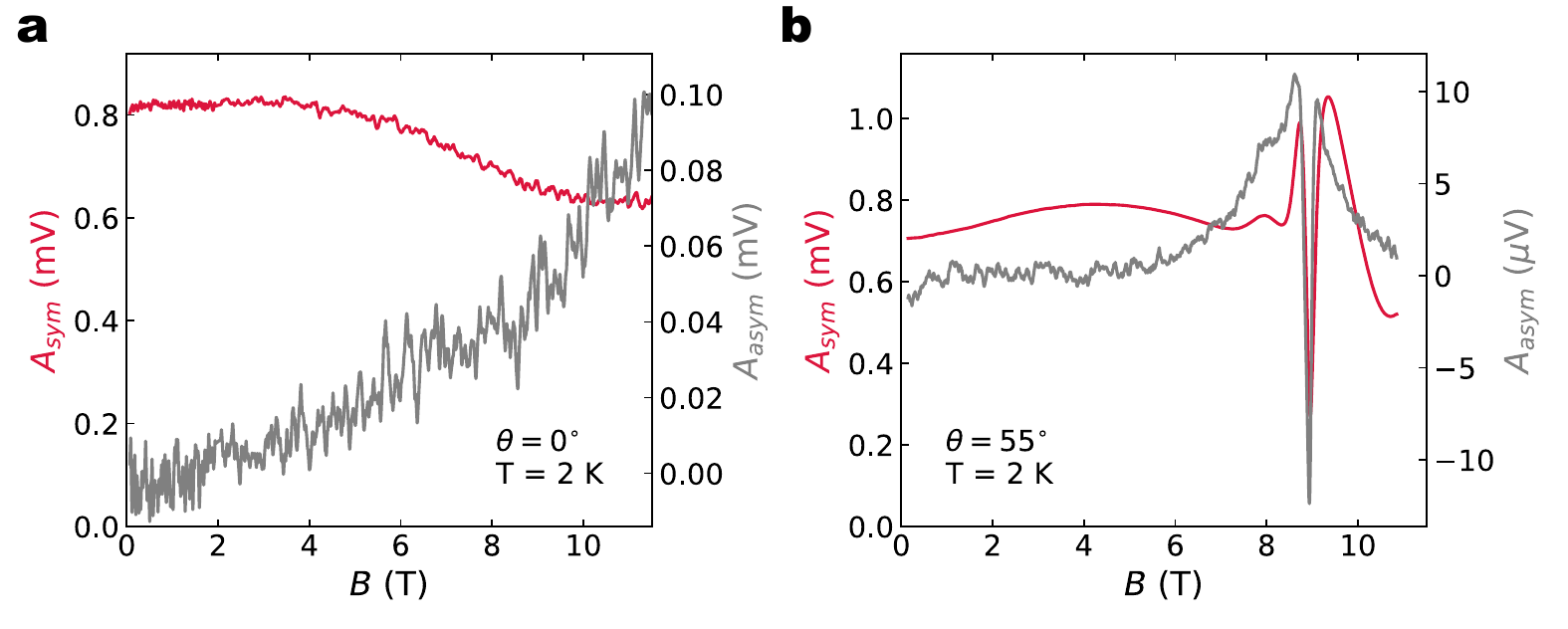}
	\end{center}
	\caption{\textbf{Reciprocal and non-reciprocal signals: }The reciprocal (symmetric: $(A(B>0) + A(B<0))/2$) and non-reciprocal (antisymmetric: $(A(B>0) - A(B<0))/2$) signals for magnetic field sweeps take with $\mathbf{B}$ along the $c$-axis (\textbf{a}) and with $\mathbf{B}$ tilted 55$^\circ$ away from the $c$-axis (\textbf{b}). In both cases, the antisymmetric signal has a distinct magnetic field dependence from the symmetric signal, further demonstrating that the measurement protocol differentiates the time-odd viscosity from time-even effects such as ultrasound attenuation. }
  \label{extfig:data-sym-asym}
\end{extfigure}

\begin{extfigure}[H]
	\begin{center}
		\includegraphics[width=.8\textwidth]{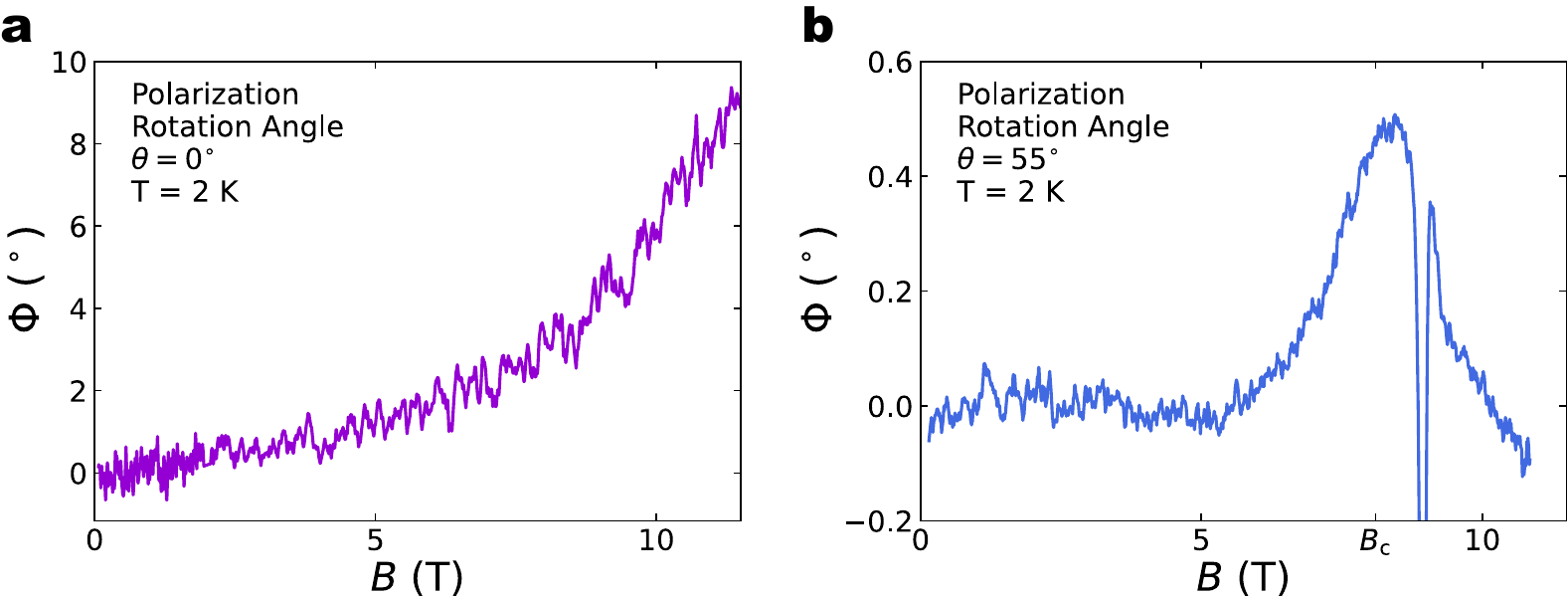}
	\end{center}
	\caption{\textbf{Antisymmetric signal in terms of polarization rotation angle: }The antisymmetric signal recast in terms of the angle through which the transverse sound wave polarization rotates as it propagates across the sample. (\textbf{a}) plots the rotation angle versus applied magnetic field for $\mathbf{B}||c$, while (\textbf{b}) shows the same information for the tilted magnetic field configuration.}
  \label{extfig:faraday-angle-field}
\end{extfigure}

\begin{extfigure}[H]
	\begin{center}
		\includegraphics[width=.8\textwidth]{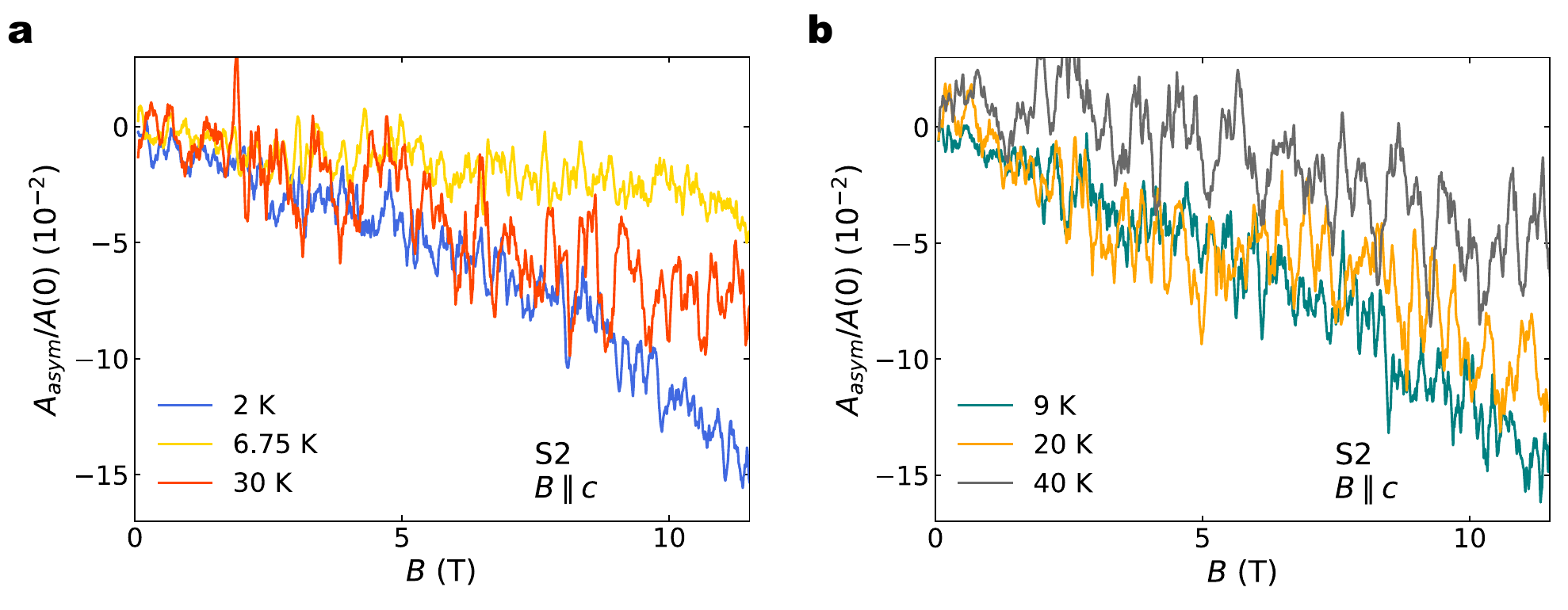}
	\end{center}
	\caption{\textbf{Temperature dependence: }To connect the Hall viscosity to the thermal Hall conductivity of \rcl we measure we measure the Faraday rotation with the magnetic field along the crystal $c$ axis as a function of temperature. Here we show the raw data used to extract the viscosity at 12 T and compare with thermal transport measurements in the main text. \textbf{a}, shows that the large antisymmetric signal measured at 2 K is reduced just below $T_{\rm{N}}$. \textbf{b}, shows that the antisymmetric signal is restored above $T_{\rm{N}}$ and decays slowly as the temperature is increased. The data are separated into two panels for visual clarity. }
  \label{extfig:high-temp-asym}
\end{extfigure}

\begin{extfigure}[H]
	\begin{center}
		\includegraphics[width=0.9\textwidth]{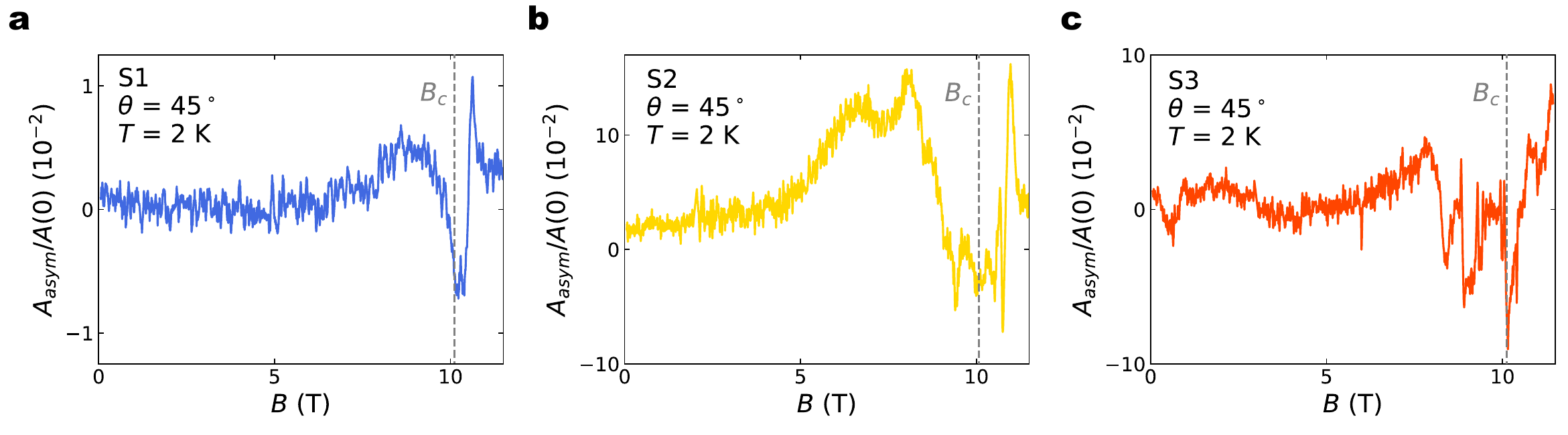}
	\end{center}
	\caption{\textbf{Sample dependence: }We show the antisymmeric in field signal for three samples of \rcl at identical temperatures (2 K) and with the magnetic field in the same orientation (tilted 45$^\circ$ away from the $c$ axis toward the $b$-axis in the honeycomb plane). The critical field, $B_c$, measured simultaneously using the compressional speed of sound is marked with a dotted vertical line. All three samples show clear evidence of Faraday rotation beginning at an onset field of $\sim$6 T. The magnitude of the antisymmetric in field signal is proportional to the thickness of the sample, which is expected because the viscosity sets the amount of rotation per unit length. There are clear differences between samples as well, notably the behavior above $B_c$ in sample 3. We ascribe these differences at least in part to the presence of structural and magnetic domains, which are especially prevalent in thicker samples.}
	\label{fig:sample-dependence}
\end{extfigure}

\begin{extfigure}[H]
	\begin{center}
		\includegraphics[width=0.9\textwidth]{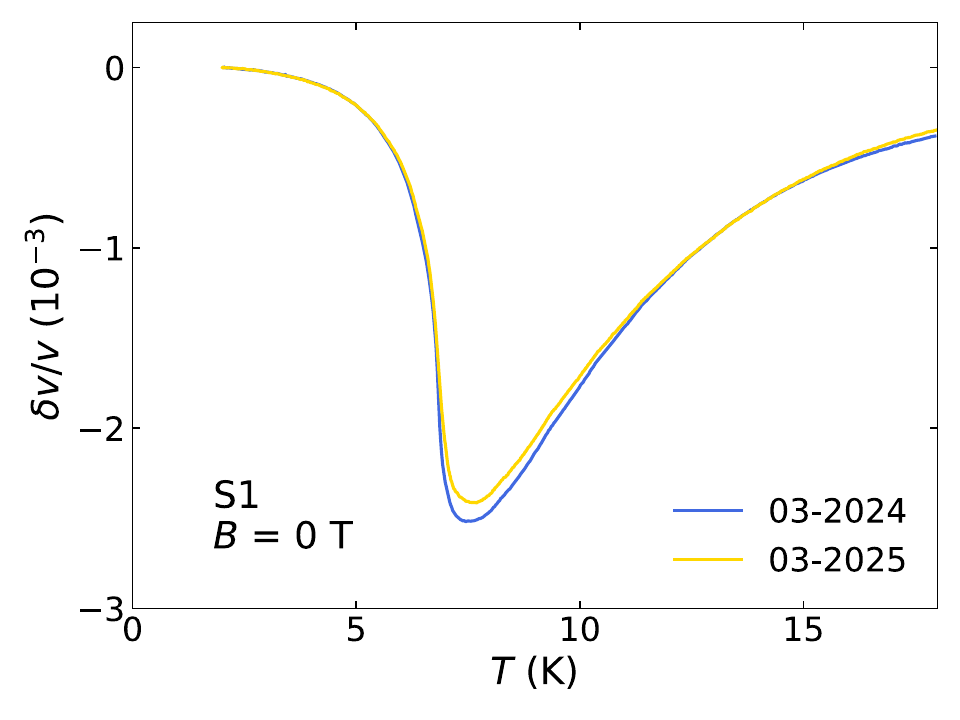}
	\end{center}
	\caption{\textbf{Effect of thermal cycling: }\rcl is known to be sensitive to thermal cycling and, for that reason, we were careful to cool and warm the sample slowly ($\sim$0.25 K/min). Here we show the speed of sound as a function of temperature before and after numerous thermal cycles for sample S1. We see no change in the sample $T_N$, evidence that the bulk properties of the sample are unchanged.}
	\label{fig:TN-time}
\end{extfigure}

\end{document}